\documentclass{jfm}
\usepackage{graphicx}
\usepackage{epstopdf, epsfig}
\usepackage{subcaption}

\usepackage{amsmath}
\usepackage{amssymb}
\usepackage{color}
\usepackage{caption}

\newcommand\bvec[1]{\mathchoice{\mbox{\boldmath$\displaystyle#1$}}
  {\mbox{\boldmath$\textstyle#1$}}
  {\mbox{\boldmath$\scriptstyle#1$}}
  {\mbox{\boldmath$\scriptscriptstyle#1$}}}
\renewcommand{\vec}[1]{\bvec{#1}}

\shorttitle{MHD stability of large scale liquid metal batteries}
\shortauthor{A. Tucs, V.  Bojarevics and K. Pericleous}

\title{MHD stability of large scale liquid metal batteries}

\author{A. Tucs,
  V.  Bojarevics \corresp{\email{V.Bojarevics@gre.ac.uk}} \and K. Pericleous }

\affiliation{University of Greenwich, Park Row, London SE10 9LS, UK}

\begin{document}

\maketitle

\begin{abstract}
The aim of this paper is to develop a stability theory and a numerical model for the three density-stratified electrically conductive liquid layers. Using regular perturbation methods to reduce the full 3d problem to the shallow layer model, the coupled wave and electric current equations are derived. The problem set-up allows the weakly non-linear velocity field action and an arbitrary vertical magnetic field. Further linearisation of the coupled equations is used for the linear stability analysis in the case of uniform vertical magnetic field. New analytical stability criteria accounting for the viscous damping are derived for particular cases of practical interest and compared to the numerical solutions for variety of materials used in the batteries. The new criteria are equally applicable to the aluminium electrolysis cell MHD stability estimates.
\end{abstract}


\section{Introduction}
Liquid Metal Batteries (LMBs) have many important characteristics for efficient practical use in combination with renewable energy sources on a national energy grid scale in the future. A relatively high voltage efficiency at high current densities of this storage technology are due to liquid-liquid electrode-electrolyte interfaces that enable high speed charge transfer, high total current capability, low ohmic losses, as well as rapid mass transport of reactants and products to and from the electrode-electrolyte interfaces by means of liquid-state diffusion \citep{Kim13}.

The liquid state of the main components necessitates consideration of the fluid dynamics in LMBs. A number of recent publications are devoted to the problem, e.g. a possibility of the Tayler instability \citep{Weber14,Herreman15}, the thermal convection \citep{Shen16}, observation of vortical flow in a LMB model \citep{Kelley14}, and a simplified model of sloshing in a three layer system \citep{Zikanov15}. The main motivation of these investigations is to prevent the possibility of direct contact between the molten metal anode and cathode that may occur due to electro-magnetically driven destabilizing interface motion. On the other hand, the controlled mixing enhances mass transport improving the cell performance, preventing the accumulation of intermetallic compounds at the electrode-electrolyte interface.

LMBs are thought to be easily scalable on the cell level due to their simple construction using the natural density stratification of the liquid layers. Large cells of several cubic meters total volume have a potential to operate at very high power value \citep{Bojarevics17}. High current densities coupled to the magnetic field (created by the currents in the cell, the supply bars and the neighbour cells) lead to significant electromagnetic forces. Such forces in stratified liquid layers with large surface areas may cause a long wave interfacial instability as it is well known in the case of Hall-Heroult cells (HHC), as first described by \citet{Sele77}. The manifestation of this instability in LMB is the subject of this paper.

In a typical HHC the electric current, of total magnitude $150-800$ kA, enters the cell from the carbon anodes, passes through the liquid electrolyte and aluminium layer, and exits via the carbon cathode blocks at the bottom of the cell. The liquid layers are relatively shallow, $4-30$ cm in depth vs $4-20$ m in horizontal dimension. The small relative depth of the layers and the small difference of the liquid densities facilitates the instability development. 

The ratio of electrical conductivities of the cell materials is another significant parameter. The liquid metal is a better conductor ($\sim10^6$ S/m) than the carbon ($\sim10^4$ S/m), while the electrolyte is about two orders of magnitude less conductive ($\sim10^2$ S/m). The significantly lower conductivity of the electrolyte means that this layer is responsible for the majority of electrical losses in the cell. Joule heating is necessary to heat the cell and to keep the metal liquid, however the total voltage drop must be as low as possible in order to achieve a better electrical efficiency. A small perturbation of the interface between liquid layers may cause a substantial redistribution of the current in the cell.

First attempts to explain the interfacial instabilities were made by \citet{Sele77}, \citet{Urata85}, \citet{Sneyd85} and \citet{Moreau86}. A more involved understanding of the physical mechanism was provided by \citet{Sneyd94}, \citet{Bojarevics94}, and \citet{Davidson98}. The mechanism is based on the standing gravity wave modification due to the electric current redistribution. The electric current density in the electrolyte increases above the wave crests, resulting in a high density horizontal current in the shallow liquid metal layer. In the presence of a vertical magnetic field the electromagnetic force excites another standing wave mode orthogonal to the initial perturbation. The new wave mode is coupled to the original mode, and the oscillation frequency is shifted. The frequency shift increases with the rise of the magnetic field until at a critical value, when the two wave frequencies coincide, an exponential growth of the amplitude indicates the onset of instability. In general, the above process is described by the following set of equations: 

\begin{equation}\label{eq:1}
\partial_{tt}H_{\vec{k}}+\mathsfbi{\omega}^2_{\vec{k}}H_{\vec{k}}=E\mathsfbi{G}_{\vec{k}\vec{k}'}H_{\vec{k}'},
\end{equation}

\noindent where $H_{\vec{k}}$ is a vector which represents the amplitudes of the original gravitational modes $\vec{k}=(k_x,k_y)$, $\mathsfbi{\omega}^2_{\vec{k}}$ is the matrix of the gravitational frequencies,  $\mathsfbi{G}_{\vec{k}\vec{k}'}$ is the interaction matrix, $E$ is the dimensionless parameter characterizing the electromagnetic forces. The mode coupling is included in  $\mathsfbi{G}_{\vec{k}\vec{k}'}$, where each column represents the Lorentz force (Fourier decomposed) in response to the gravitational wave modes. These coupled equations represent an eigenvalue problem for the square of the new complex frequencies $\mu$ ($H_{\vec{k}}\sim e^{\mu t}$). The matrix $\mathsfbi{G}_{\vec{k}\vec{k}'}$ is real anti-symmetric, and in a general case the eigenvalues are shifted increasing the magnetic field \citep{Bojarevics94}. Onset of the instability starts at a critical value of $E$ at which the exponentially growing part of the complex eigenvalue $\mu$ appears. See also the paper by \citet{Antille02} where the numerical eigenvalue solution is analysed when approaching the  instability threshold. A key point noted in the papers, is that the dominant contribution to the perturbed Lorentz force arises from the interaction between a horizontal current in the aluminium layer and the vertical component of the background magnetic field. 

\citet{Davidson98} derived a simple mechanical analogue which captures the basic features of the instability. The liquid aluminium layer is represented by a compound pendulum that consists of a large flat aluminium plate attached to a top surface by a light, rigid strut. The strut is pivoted at its top end so that the plate is free to swing along two horizontal axes $x$ and $y$. The fluid system of infinite motion freedom is reduced to only two degrees of freedom. \citet{Zikanov15} constructed a similar mechanical model for instability description in the LMB taking into account an additional top liquid metal layer. The metal layers of the battery are represented by solid metal slabs rigidly attached to weightless rigid struts pivoted at the top. The free oscillations of the slabs imitate the sloshing motion of the liquid layers. The slabs are separated from each other by a layer of a poorly conducting electrolyte. Two destabilization mechanisms were considered: 1) interaction of a purely vertical magnetic field and horizontal currents, similar to HHC, 2) interaction between the current perturbations and the azimuthal self-magnetic field from the total vertical current. The first mechanism will occur in real batteries if a sufficiently strong vertical magnetic field is present due to the presence of external current supply. The batteries of a square or a circular horizontal cross section will be always unstable if even a small field is present. The second mechanism appears to be more challenging since the azimuthal magnetic field, unlike the vertical magnetic field, cannot be reduced via optimization of the current supply lines unless they cross the liquid layer \citep{Weber14}. The existence of the second instability type was predicted by \citet{Munger08} for HHC case, yet needs more clarification for the LMB case. The approach developed by \citet{Davidson98} and \citet{Zikanov15} is purely mechanical. However, the principal physical mechanism could be valid, due to the fact that sloshing motions generated in the shallow liquid layers are inherently large scale, and so their qualitative behaviour can be approximately described using the coupled pair of long wave modes approach. 

More realistic fluid dynamic description can be achieved starting from the full set of Navier-Stokes equations by means of the shallow layer approximation and a systematic derivation of a set of coupled wave equations governing the three fluid layers.  The hydrodynamic coupling is realised by pressure continuity at the common interfaces. The continuity of the electric potential and the supplied electric current will introduce the electromagnetic coupling of waves. In the following sections the linear stability of coupled modes will be investigated in the presence of a purely vertical magnetic field, accounting for the continuous electric current in the LMB model. The role of dissipation rate will be analysed using both analytical tools and numerical solutions. Analytical criteria for the cell stability will be established using approximations suitable for a practical cell design, including the solid bottom and top friction effects on the shallow layers.
 
\section{Interfacial dynamics}\label{sec:rules_submission}

\begin{figure}
  \centerline{\includegraphics[width=0.55\textwidth]{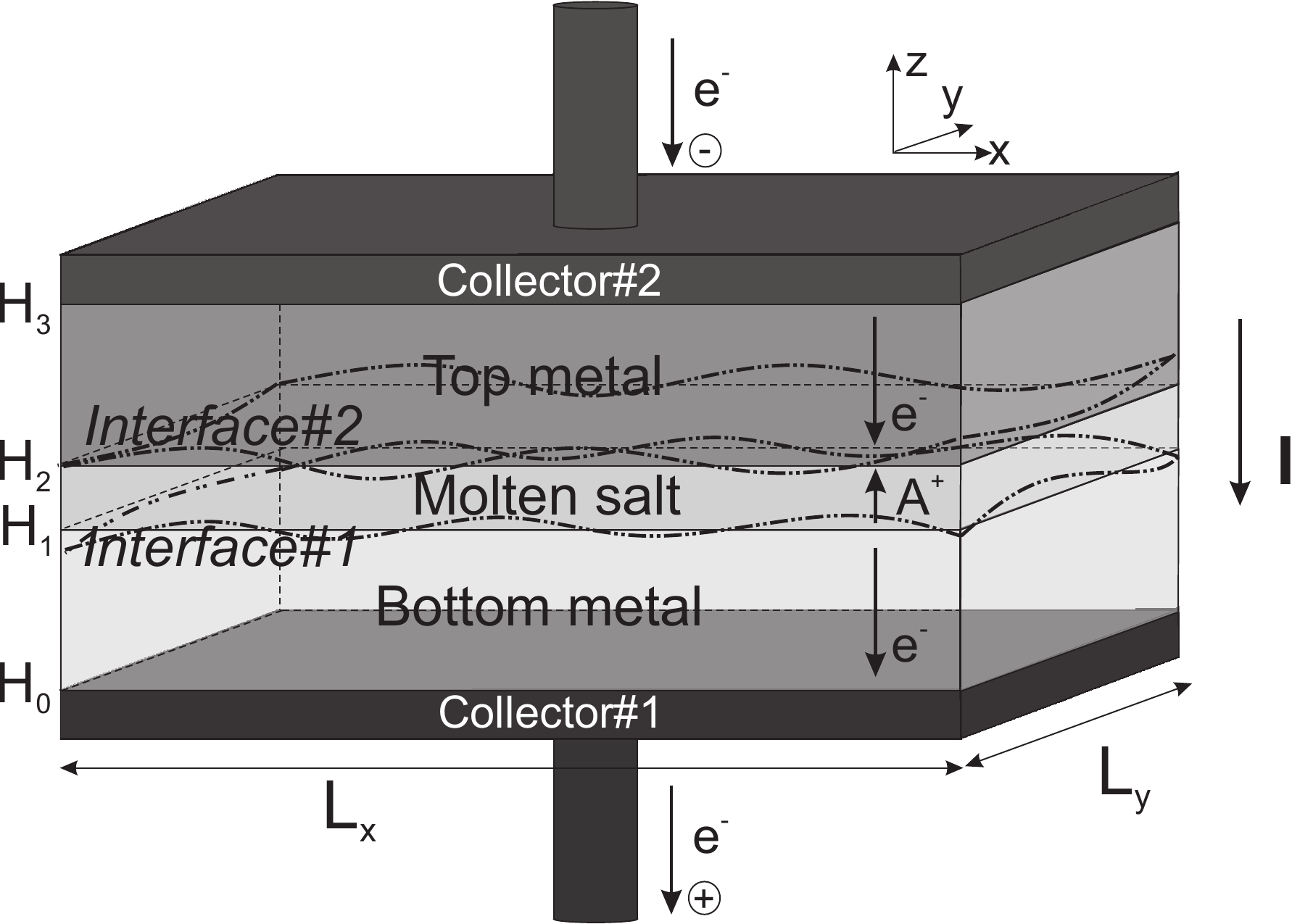}}
  \caption{3 layer liquid metal model under consideration.}
\label{fig:1}
\end{figure}

Hydrodynamics of the three density stratified electrically conductive liquid layer system, schematically represented in figure \ref{fig:1}, in the presence of electro-magnetic fields, are described by the following equations

\begin{equation}\label{eq:2}
\rho\partial_tu_i+\rho u_j\partial_ju_i=-\partial_i(p+\rho gz)+\partial_j\rho\nu(\partial_ju_i+\partial_iu_j)+f_i,
\end{equation}

\begin{equation}\label{eq:3}
\partial_iu_i=0,
\end{equation}

\noindent where the indices $i,j=1,2,3$ correspond to the coordinates $(x,y,z)$, the velocity components are given as $(u_1,u_2,u_3)$, the summation over repeated indices is implied, $\rho$ represents density, $\nu$ - effective viscosity, $g$ - the gravitational acceleration, $p$ - pressure and the vector of Lorentz force $\vec{f}=\vec{j}\times\vec{B}$ is computed in each of the 3 layers, $\vec{j}$ is the current density and $\vec{B}$ the magnetic field. In this paper the horizontal dimensions of the cell are assumed to be much larger compared to the vertical depth, so that the description can be based on a systematically derived shallow layer approximation. The velocity components in each layer can be represented as an expansion in a small aspect ratio parameter $\delta=\max h/\min L$, where $h$ is a typical depth, for instance the unperturbed metal layer, and $L$ is the characteristic horizontal dimension (width of the cell):

\begin{equation}\label{eq:4}
u_{i}=u_{0i}(x,y,t)+\delta u_{1i}(x,y,\overline{z},t)+\textit{O}(\delta^2), \qquad i=1,2,
\end{equation}

\begin{equation}\label{eq:6}
u_{3}=\delta u_{13}(x,y,\overline{z},t)+\textit{O}(\delta^2),
\end{equation}

\noindent where a stretched vertical coordinate $\overline{z}=z/\delta$ is introduced. The $u_{3}$ expansion starts with the $\delta$-order due to (\ref{eq:3}). If all three components of the electromagnetic force density are of the same order of magnitude: $f_x\sim f_y\sim f_z$, and the horizontal pressure gradient components are of the same order as the corresponding force components: $\partial_{i}p\approx f_{i},$ then the vertical component of the gradient $\partial_3p\sim -\rho g\gg f_z$. According to these estimates, the leading horizontal $(i=1,2)$ components of (\ref{eq:2}) are

\begin{equation}\label{eq:611}
\rho\partial_tu_{0i}+\rho u_{0j}\partial_ju_{0i}=-\partial_ip+\delta^{-1}\partial_{\overline{z}}\rho\nu\partial_{\overline{z}}(u_{0i}+\delta u_{1i})+f_i.
\end{equation}

\noindent The vertical component of the equation (\ref{eq:2}) gives the leading order terms as:

\begin{equation}\label{eq:612}
-\delta^{-1}\partial_{\overline{z}}(p+\rho g\delta\overline{z})+f_{z}=0.
\end{equation}

\noindent The hydrostatic pressure in the liquid layers adjacent to the interface $H_1(x,y,t)$, see the figure \ref{fig:1}, can be expressed by

\begin{equation}\label{eq:7}
p_m^{(1)}(x,y,z,t)=p_p^{(1)}(x,y,t)-\rho_mg(z-H_1)+\delta\int_{H_1(x,y,t)}^zf_{zm}dz,
\end{equation}

\noindent where the index $m=1,2$ stands for the layer number and $p_p^{(1)}$ is the reference pressure at the moving interface $z=H_1(x,y,t)$. Similarly the same pressure can be referenced to the interface $H_2(x,y,t)$, the corresponding pressure given by

\begin{equation}\label{eq:8}
p_n^{(2)}(x,y,z,t)=p_{p}^{(2)}(x,y,t)-\rho_ng(z-H_2)+\delta\int_{H_2(x,y,t)}^{z}f_{zn}dz,
\end{equation}

\noindent where in this particular case $n=2,3$. The respective horizontal gradients of the pressure required in the horizontal momentum equation (\ref{eq:611}) are:

\begin{equation}\label{eq:9}
\partial_ip_m^{(1)}=\partial_ip_{p}^{(1)}+\rho_mg\partial_iH_1+\delta\left[\int_{H_1}^z\partial_if_{z}dz-f_z(H_1)\partial_i H_1\right],
\end{equation}

\begin{equation}\label{eq:10}
\partial_ip_n^{(2)}=\partial_ip_{p}^{(2)}+\rho_ng\partial_iH_2+\delta\left[\int_{H_2}^z\partial_if_{z}dz-f_z(H_2)\partial_i H_2\right].
\end{equation}

The next step is to introduce the depth averaging within each layer. The depth averaging for horizontal velocity components is performed in the following way:

\begin{equation}\label{eq:11}
U_{ik}=h_k^{-1}\int_{H_{k-1}}^{H_k}u_{ik}(x,y,z)dz,
\end{equation}

\noindent where $k=1,2,3$ is the layer number (no summation over $k$) and $h_k(x,y,t)=H_{k}-H_{k-1}$ is the local variable depth, see figure \ref{fig:1}. The depth averaging can be applied to the continuity equation (\ref{eq:3}):

\begin{eqnarray}\label{eq:12}
h_k^{-1}\int_{H_{k-1}}^{H_k}(\partial_iu_i+\partial_{3}u_3)dz & = & h_k^{-1}\left[ \astrut \partial_i\int_{H_{k-1}}^{H_k}u_idz-u_i(H_k)\partial_iH_k+u_i(H_{k-1})\partial_iH_{k-1}\right. \nonumber\\
&& \left.\mbox{} + u_3(H_k)-u_3(H_{k-1})\astrut\right]=0 ,
\label{eq43}
\end{eqnarray}

\noindent where $i=1,2$. The vertical velocity $u_3$ at the $z=H_k(x,y,t)$ is given by the kinematic condition, stating that the interface moves with the local velocity:

\begin{equation}\label{eq:13}
u_3(H_k)=\partial_tH_k+u_i(H_k)\partial_iH_k.
\end{equation}

\noindent Substituting (\ref{eq:13}) into (\ref{eq:12}) leads to

\begin{equation}\label{eq:14}
\partial_th_k+\partial_i(U_{ik}h_k)=0.
\end{equation}

\noindent The last equation can be linearised if an additional approximation of a small wave amplitude is introduced: $h_k(x,y,t)=h_{0k}+\varepsilon h_k'(x,y,t)$ for the layer thickness or equivalently $H_m(x,y,t)=H_{0m}+\varepsilon\zeta_m(x,y,t)$ for the interface position, where the additional small parameter $\varepsilon=\max A/h$ is introduced. $A$ is a typical wave amplitude and $h_{0k}$, $H_{0m}$ are the unperturbed values, $\zeta_m$ are the interfacial perturbations. For each particular layer the depth average horizontal velocity divergence can be expressed as:

\begin{equation}\label{eq:15}
\partial_iU_{i1}=-\frac{\varepsilon}{h_{01}}[\partial_t\zeta_1+\partial_i(U_{i1}\zeta_1)],
\end{equation}

\begin{equation}\label{eq:16}
\partial_iU_{i2}=-\frac{\varepsilon}{h_{02}}\lbrace\partial_t(\zeta_2-\zeta_1)+\partial_i[U_{i2}(\zeta_2-\zeta_1)]\rbrace,
\end{equation}

\begin{equation}\label{eq:17}
\partial_iU_{i3}=\frac{\varepsilon}{h_{03}}[\partial_t\zeta_2+\partial_i(U_{i3}\zeta_2)].
\end{equation}

\noindent The volume conservation requires:

\begin{equation}\label{eq:18}
h_1+h_2+h_3=H_3-H_0.
\end{equation}

In order to estimate the leading order of terms in the depth averaged momentum equations (\ref{eq:611}) dimensionless variables of order $\textit{O}(1)$ are introduced using the following scaling: $L$ - for the coordinates $x,y$; $h$ for a typical layer thickness; $\varepsilon\sqrt{gh}$ - for the wave velocity; $L/\sqrt{gh}$ - for the time; $\rho_kgh$ for the pressure ($k=1,2,3$). For typical geometries considered in this paper $\delta=h/L\approx$ 0.2 m/8 m = $0.0250\ll1$, whereas, $\varepsilon=A/h\approx$ 0.005 m/0.2 m = $0.0250\ll1$. 

The horizontal pressure gradient from the expressions (\ref{eq:9}) and (\ref{eq:10}), neglecting terms of the $\delta$ and higher order, can be substituted in the depth averaged, nondimensionalized horizontal momentum equation (\ref{eq:611}). For the layers adjacent to the lower interface $H_1(x,y,t)$ the respective momentum equations are

\begin{equation}\label{eq:19}
\partial_tU_{i1}+\varepsilon U_{j1}\partial_jU_{i1}=-\varepsilon^{-1}\rho_1^{-1}\partial_ip_{p}^{(1)}-g\partial_i\zeta_1-k_{f1}U_{i1}+E_1F_{i1},
\end{equation}

\begin{equation}\label{eq:20}
\partial_tU_{i2}+\varepsilon U_{j2}\partial_jU_{i2}=-\varepsilon^{-1}\rho_2^{-1}\partial_ip_{p}^{(1)}-g\partial_i\zeta_1-k_{f2}U_{i2}+E_2F_{i2},
\end{equation}

\noindent where the depth averaged force $F_i$ is defined similarly to (\ref{eq:11}). For the upper interface $H_2(x,y,t)$ the respective equations are:

\begin{equation}\label{eq:21}
\partial_tU_{i2}+\varepsilon U_{j2}\partial_jU_{i2}=-\varepsilon^{-1}\rho_2^{-1}\partial_ip_{p}^{(2)}-g\partial_i\zeta_2-k_{f2}U_{i2}+E_2F_{i2},
\end{equation}

\begin{equation}\label{eq:22}
\partial_tU_{i3}+\varepsilon U_{j3}\partial_jU_{i3}=-\varepsilon^{-1}\rho_3^{-1}\partial_ip_{p}^{(2)}-g\partial_i\zeta_2-k_{f3}U_{i3}+E_3F_{i3}.
\end{equation}

\noindent The equations (\ref{eq:20}) and (\ref{eq:21}) formally give the connection between the reference pressures $p_{p}^{(1)}$ and $p_{p}^{(2)}$ defined on the two interfaces, however being valid in the same fluid layer $k=2$ (electrolyte). The alternative representations are required for the wave equation derivation. After the integration over depth the dissipative terms in (\ref{eq:611}) are replaced by empirical expressions used for the shallow layer approximation \citep{Moreau84, Rodi87} using a linear in velocity friction law with the coefficients $k_{fk}$. The electromagnetic interaction parameter (the ratio of electromagnetic force to the gravity force perturbation) is introduced as $E_k=IB_0/(L^2\rho_kg\varepsilon\delta)$, where $I$ is the total electric current, $B_0$ is a typical magnitude of magnetic field. The corresponding magnitude of $E$ can be estimated, using typical values for $I=10^5$ A, $B_0=10^{-3}$ T, $L=8$ m (width of cell), $\rho=1.6\times10^3$ kg m$^{-3}$ (liquid magnesium for the top metal), $g=9.8$ m s$^{-2}$, $\varepsilon=\delta=0.025$: $E=0.32=\textit{O}(1)$. The electromagnetic term is of the same order of magnitude as the leading terms, while the nonlinear wave motion terms are of lower order ($\sim\varepsilon$) and will be neglected later in the linear theory, but retained for a numerical solution.

The wave equations for the coupled interfaces can be derived following the procedure described in \citet{Bojarevics92}:

\begin{enumerate}
\item take time derivative of the non-dimensional linearised equations (\ref{eq:15}),(\ref{eq:16}):

\begin{equation}\label{eq:155}
\partial_{it}U_{i1}=-\frac{\varepsilon}{h_{01}}\partial_{tt}\zeta_1,
\end{equation}

\begin{equation}\label{eq:166}
\partial_{it}U_{i2}=-\frac{\varepsilon}{h_{02}}\partial_{tt}(\zeta_2-\zeta_1),
\end{equation}

\item substitute (\ref{eq:155}), (\ref{eq:166}) into the horizontal divergence of (\ref{eq:19}),(\ref{eq:20}),
\item take the difference of the resulting equations.
\end{enumerate}

\noindent This procedure eliminates the common unknown pressure $p_{p}^{(1)}$ on the interface $\zeta_1$:

\begin{eqnarray}\label{eq:27}
&&\alpha_1\partial_{tt}\zeta_1+k_{fe1}\partial_{t}\zeta_1-\frac{\rho_2}{h_{02}}\partial_{tt}\zeta_2-\frac{\rho_2k_{f2}}{h_{02}}\partial_{t}\zeta_2 \nonumber\\
&&=R_1\partial_{jj}\zeta_1-\rho_1E_1\partial_iF_{i1}+\rho_2E_2\partial_iF_{i2}\nonumber\\
&&+\varepsilon[\rho_1\partial_{j}(U_{k1}\partial_{k}U_{j1})-\rho_2\partial_{j}(U_{k2}\partial_{k}U_{j2})].
\end{eqnarray}

\noindent The corresponding boundary conditions for the normal velocity $u_n=0$ at the side walls can be obtained by taking the difference between (\ref{eq:19}) and (\ref{eq:20}) to eliminate the common pressure at the interface $\zeta_1$: 

\begin{equation}\label{eq:28}
\partial_n\zeta_1=(\rho_1E_1F_{n1}-\rho_2E_2F_{n2})/R_1.
\end{equation}

\noindent In a similar manner the wave equation for the upper interface $\zeta_2$ can be obtained:

\begin{eqnarray}\label{eq:29}
&&\alpha_2\partial_{tt}\zeta_2+k_{fe2}\partial_{t}\zeta_2-\frac{\rho_2}{h_{02}}\partial_{tt}\zeta_1-\frac{\rho_2k_{f2}}{h_{02}}\partial_{t}\zeta_1\nonumber\\
&&=R_2\partial_{jj}\zeta_2-\rho_2E_2\partial_iF_{i2}+\rho_3E_3\partial_iF_{i3}\nonumber\\
&&+\varepsilon[\rho_2\partial_{j}(U_{k2}\partial_{k}U_{j2})-\rho_3\partial_{j}(U_{k3}\partial_{k}U_{j3})],
\end{eqnarray}

\noindent and the corresponding boundary conditions are

\begin{equation}\label{eq:30}
\partial_n\zeta_2=(\rho_2E_2F_{n2}-\rho_3E_3F_{n3})/R_2.
\end{equation}

\noindent The new constants introduced in the above equations are defined as:

\begin{equation}\label{eq:471}
\alpha_1=\frac{\rho_1}{h_{01}}+\frac{\rho_2}{h_{02}},\quad \alpha_2=\frac{\rho_2}{h_{02}}+\frac{\rho_3}{h_{03}},
\end{equation}

\begin{equation}\label{eq:472}
k_{fe1}=\frac{\rho_1k_{f1}}{h_{01}}+\frac{\rho_2k_{f2}}{h_{02}},\quad k_{fe2}=\frac{\rho_3k_{f3}}{h_{03}}+\frac{\rho_2k_{f2}}{h_{02}},
\end{equation}

\begin{equation}\label{eq:473}
R_1=(\rho_1-\rho_2)g,\quad R_2=(\rho_2-\rho_3)g.
\end{equation}

\noindent Note, that in (\ref{eq:27}) and (\ref{eq:29}) summation over repeated indices is limited to the two horizontal dimensions. As it can be seen from (\ref{eq:27}) and (\ref{eq:29}), both interfaces can not be considered independently due to the presence of coupling terms. In the following we will  assume that the interfacial friction $k_{f2}$ is negligible. The set of the equations in the electrically nonconductive limit and in the absence of viscous dissipation is in correspondence with the set of the equations obtained by \citet{Robino01} derived for the dynamics of internal solitary waves in stratified 3 layer ocean.  The aluminium electrolysis cell MHD wave model can be recovered if $\zeta_2=0$ in (\ref{eq:27}).

\section{Electric current flow}\label{sec:types_paper}

For energy storage and supply the LMBs must operate in two regimes: charge and discharge, resulting in the current flowing (upwards or downwards). In this paper only the charging process is considered due to the physical symmetry of both operational regimes. The current flow in the layered structure is illustrated in figure \ref{fig:1}. Similarly to \citet{Davidson98}, we assume that the characteristic time-scale for the wave motion is much larger than the diffusion time of the magnetic field to satisfy the low magnetic Reynolds number approximation, leading to $Rm=\mu\sigma Uh\ll1$, where $\mu$ is the magnetic permeability, $\sigma$ the electrical conductivity, $U$ is a typical velocity. Using typical values for the liquid metal $\mu=4\pi\times10^{-7}$ H m$^{-1}$, $\sigma=3.65\times10^6$ S m$^{-1}$, $U=0.01$ m s$^{-1}$, $h=0.1$ m the estimated value of $Rm\approx0.004$. A similar estimate can be obtained for the wave motion magnetically induced electric current ration to the basic current density: $\sigma UB/(I/L^2)\sim\sigma\varepsilon\sqrt{gh}B_0/(I/L^2)\sim0.001$. 

In the low $Rm$ approximation, when the flow effect is neglected, the electric current can be expressed as

\begin{equation}\label{eq:310}
\vec{j}_k=-\sigma_k\vec{\nabla}\varphi_k
\end{equation}

\noindent and is described by a set of coupled Laplace equations for the electric potential $\varphi_k(x,y,z)$: 

\begin{equation}\label{eq:31}
\partial_{ii}\varphi_k=0,
\end{equation}

\noindent where $k=1,2,3$ corresponds to the layer number. The continuity conditions for the electric potential and the normal current component $\vec{j}\cdot\vec{n}$ at the interfaces $z=H_m$ $(m=1,2)$ are

\begin{equation}\label{eq:32}
\varphi_m=\varphi_{m+1},
\end{equation}

\begin{equation}\label{eq:33}
\sigma_{m+1}\partial_n\varphi_{m+1}=\sigma_m\partial_n\varphi_m.
\end{equation}

\noindent The normal derivatives at the deformed interfaces are defined as (assuming the summation over the repeated index $i$ only):

\begin{equation}\label{eq:34}
\partial_n\varphi_k=\frac{\partial_{z}\varphi_k-\partial_iH_k\partial_i\varphi_k}{(1+\partial_iH_k\partial_iH_k)^{1/2}}.
\end{equation}

\noindent With (\ref{eq:34}) the current continuity (\ref{eq:33}) at the interfaces $H_1$ and $H_2$ can be written explicitly in the nondimensional form in order to estimate the leading order terms:

\begin{equation}\label{eq:35}
s_1\delta^2(\partial_{\overline{z}}\varphi_2-\varepsilon\delta^2\partial_i\overline{\zeta}_1\partial_i\varphi_2)=\partial_{\overline{z}}\varphi_{1}-\varepsilon\delta^2\partial_i\overline{\zeta}_{1}\partial_i\varphi_{1},
\end{equation}

\begin{equation}\label{eq:36}
\partial_{\overline{z}}\varphi_{3}-\varepsilon\delta^2\partial_i\overline{\zeta}_{2}\partial_i\varphi_{3}=s_3\delta^2(\partial_{\overline{z}}\varphi_2-\varepsilon\delta^2\partial_i\overline{\zeta}_2\partial_i\varphi_2),
\end{equation}

\noindent where the four orders of magnitude difference in the electrical conductivities permit to define $\sigma_2/\sigma_1=s_1\delta^2$, $\sigma_2/\sigma_3=s_3\delta^2$ and the stretched $\overline{\zeta}_i=\zeta_i/\delta$. These definitions allow us to compare numerically the electrical conductivities in the poorly conducting electrolyte relative to the well conducting liquid metals, and the effect of the small depth ($\sim\delta$) of the layers. The side walls of the domain are considered to be electrically insulating:

\begin{equation}\label{eq:37}
(\partial_n\varphi_k)_{x=0,L_x;y=0,L_y}=0.
\end{equation}

\noindent In this paper we assume, that the applied current distributions at the top and the bottom are uniform, equal and the interfacial perturbation do not influence the electric current distributions in the collectors:

\begin{equation}\label{eq:38}
(j)_{\overline{z}=\overline{H}_0}=(j)_{\overline{z}=\overline{H}_3}=-j,
\end{equation}

\noindent where $\overline{H}_k=H_k/\delta$. In principle, $j(x,y,t)$ could be used, however requiring an external circuit solution. 

The set of Laplace equations (\ref{eq:31}) can be rewritten in a nondimensional form

\begin{equation}\label{eq:40}
\delta^2\partial_{ii}\varphi_k+\partial_{\overline{z}\overline{z}}\varphi_k=0.
\end{equation}

\noindent The shallow layer approximation requires that the potential is expanded in terms of the parameter $\delta$:

\begin{equation}\label{eq:41}
\varphi_k(x,y,\overline{z},t)=\varphi_{0k}+\delta\varphi_{1k}+\delta^2\varphi_{2k}+\textit{O}(\delta^3),
\end{equation}

\noindent where the expansion terms are expressed in a similar manner as in \citet{Bojarevics94}:

\begin{equation}\label{eq:42}
\varphi_{0k}=(a_k+\varepsilon A_k)\overline{z}+(b_k+\varepsilon B_k),
\end{equation}

\begin{equation}\label{eq:43}
\varphi_{1k}=(c_k+\varepsilon C_k)\overline{z}+(d_k+\varepsilon D_k),
\end{equation}

\begin{equation}\label{eq:44}
\varphi_{2k}=(e_k+\varepsilon E_k)\overline{z}+(g_k+\varepsilon G_k)-\frac{1}{6}\overline{z}^3\partial_{ii}(a_k+\varepsilon A_k)-\frac{1}{2}\overline{z}^2\partial_{ii}(b_k+\varepsilon B_k),
\end{equation}

\noindent where $a$, $b$, $c$, $d$, $e$, $g$  are the coordinate $x$ and $y$ dependent functions that correspond to the unperturbed interfaces. The functions $A$, $B$, $C$, $D$, $E$, $G$ are $x$, $y$ and time $t$ dependent, corresponding to the perturbed interfaces. Taking into account the previously described boundary conditions and neglecting the higher order terms, the unknown coefficients can be determined as shown in the Appendix A by equalising the similar order of magnitude terms.

Finally, the resulting set of the equations governing the electric current distribution in the system is obtained by introducing the perturbed potentials in both metal layers: $\Phi_{1}=\varepsilon B_{1}$, $\Phi_{3}=\varepsilon B_{3}$. For application in the LMB the dimensional equations for the electric potential perturbations are linearly correlated to the respective interface perturbations:

\begin{equation}\label{eq:45}
h_{01}h_{02}\partial_{kk}\Phi_1-\sigma_{e,1}\Phi_1=\frac{j}{\sigma_1}(\zeta_2-\zeta_1),
\end{equation}

\begin{equation}\label{eq:46}
h_{02}h_{03}\partial_{kk}\Phi_3-\sigma_{e,2}\Phi_3=-\frac{j}{\sigma_3}(\zeta_2-\zeta_1).
\end{equation}

\noindent where

\begin{equation}\label{eq:465}
\sigma_{e,1}=\frac{\sigma_2}{\sigma_1}\left(1+\frac{\sigma_1}{\sigma_3}\frac{h_{01}}{h_{03}}\right),
\end{equation}

\begin{equation}\label{eq:466}
\sigma_{e,2}=\frac{\sigma_2}{\sigma_3}\left(1+\frac{\sigma_3}{\sigma_1}\frac{h_{03}}{h_{01}}\right),
\end{equation}

\noindent where $h_{0k}$ is the unperturbed layer thickness. The current distribution in the electrolyte is almost purely vertical due to fact that $\sigma_2\ll\sigma_1\sim\sigma_3$ (similarly to \citet{Zikanov17}). According to (\ref{eq:45}) and (\ref{eq:46}), the current flow is perturbed by the electrolyte thickness perturbations. Finally, the corresponding dimensional current components can be expressed as

\begin{equation}\label{eq:461}
\vec{j}_1=-\sigma_1\left(\partial_{x}\Phi_{1},\partial_{y}\Phi_{1},\frac{j}{\sigma_1}+(H_0-z)\partial_{ii}\Phi_{1}\right),
\end{equation}

\begin{equation}\label{eq:464}
\vec{j}_2=-\sigma_2\left(0,0,\left(1-\frac{h_2-h_{02}}{h_{02}}\right)\frac{j}{\sigma_2}+\frac{\Phi_{3}-\Phi_{1}}{h_{02}}\right),
\end{equation}

\begin{equation}\label{eq:462}
\vec{j}_3=-\sigma_3\left(\partial_{x}\Phi_{3},\partial_{y}\Phi_{3},\frac{j}{\sigma_3}+(H_3-z)\partial_{ii}\Phi_3\right).
\end{equation}

In the following section it will be shown that some of these perturbations may become electromagnetically coupled due to the presence of magnetic field and may lead to an instability resulting in a short circuit state at the extreme case.

\section{Linear stability analysis}\label{sec:filetypes}

\subsection{Coupled 3 layer problem}

The wave equations (\ref{eq:27})-(\ref{eq:30}) can be linearised neglecting the $\varepsilon$-order nonlinear terms and assuming a given magnetic field distribution, which can be expanded in terms of the small parameter $\delta$: $\vec{B}(x,y,z)=\vec{B}^0(x,y)+\delta \vec{B}^1(x,y,z)+\textit{O}(\delta^2)$. This expansion follows the same principle as the electric current representation $\vec{j}(x,y,z)=\vec{j}^0(x,y)+\delta \vec{j}^1(x,y,z)+\textit{O}(\delta^2)$. At this stage we assume that the leading part of the magnetic field is caused by external sources (connectors, supply lines, neighbouring batteries etc.), therefore $\vec{\nabla}\times\vec{B}^0=0$ in the liquid zone. This allows us to neglect the EM force components resulting from the horizontal magnetic field interaction with the vertical unperturbed current: $-j^0_zB^0_y\vec{e}_x$, $j^0_zB^0_x\vec{e}_y$ due to the zero contribution to the horizontal divergence term in (\ref{eq:27}), (\ref{eq:28}): $j^0_z(\partial_yB^0_x-\partial_xB^0_y)\vec{e}_z=0$. The leading horizontal force components contain the horizontal electric current and the vertical magnetic field: $j^1_yB^0_z\vec{e}_x$, $-j^1_xB^0_z\vec{e}_y$,  noting that $j_z^1$ is $\delta$-order lower than the horizontal perturbation current. This confirms to the assumptions made in the Introduction and implied in the previous studies on MHD stability of HHC, that the magnetic field is purely-vertical $\vec{B}=\vec{B}^0=B_z^{0}(x,y)\vec{e}_z$, and it is caused by external sources. 

The set of the wave equations for this case has the following form, after assuming that the friction at the electrolyte top and bottom is negligible in comparison to the friction at the solid top and bottom:

\begin{eqnarray}\label{eq:47}
\alpha_1\partial_{tt}\zeta_1 &+& k_{f1}\partial_{t}\zeta_1-\frac{\rho_2}{h_2}\partial_{tt}\zeta_2 \nonumber\\
&&=R_1\partial_{jj}\zeta_1+\sigma_1(\partial_y\Phi_1\partial_xB_z^{0}-\partial_x\Phi_1\partial_yB_z^{0}),
\end{eqnarray}

\begin{eqnarray}\label{eq:48}
\alpha_2\partial_{tt}\zeta_2 &+&k_{f3}\partial_{t}\zeta_2-\frac{\rho_2}{h_2}\partial_{tt}\zeta_1\nonumber\\
&&=R_2\partial_{jj}\zeta_2+\sigma_3(\partial_x\Phi_3\partial_yB_z^{0}-\partial_y\Phi_3\partial_xB_z^{0}),
\end{eqnarray}

\noindent with the boundary conditions at the side-walls:

\begin{equation}\label{eq:49}
R_1\partial_n\zeta_1-B_z^{0}\sigma_1(n_y\partial_x\Phi_1-n_x\partial_y\Phi_1)=0,
\end{equation}

\begin{equation}\label{eq:50}
R_2\partial_n\zeta_2-B_z^{0}\sigma_3(n_x\partial_y\Phi_3-n_y\partial_x\Phi_3)=0.
\end{equation}

The electric potential distribution is governed by the set of equations (\ref{eq:45}), (\ref{eq:46}). Note, that the coefficients on the left hand side of (\ref{eq:45}) and (\ref{eq:46}) contain only the constant parts of the layer thickness. 

The problem can be rewritten in a weak form by means of integrating the equations on the horizontal interface $\Gamma\in[0,L_x;0,L_y]$ against a set of regular test functions (see the integral representation in Appendix B). The solution can be constructed in a Sobolev space $W^{1,2}(\Gamma)$, so that $\zeta$ and $\Phi$ are satisfying the corresponding equations for all test-functions $\psi$ and $q$ that belong to $W^{1,2}(\Gamma)$.  The following set of functions is introduced:

\begin{equation}\label{eq:54}
\Lambda=\left\lbrace\frac{2}{\sqrt{L_xL_y}}\epsilon_{\vec{k}}\cos(k_xx)\cos(k_yy);k_x=\frac{m\upi}{L_x},k_x=\frac{n\upi}{L_y}; m,n\in N\right\rbrace,
\end{equation}

\[ \epsilon_{\vec{k}} = \left\{ \begin{array}{lll}
         1 & \mbox{if $k_x,k_y\neq 0$},\\
        1/\sqrt{2} & \mbox{if $k_x$ or $k_y$ $= 0,k_x\neq k_y$},\\
        1/2 & \mbox{if $k_x=k_y=0$},\end{array} \right. \]

\noindent where $\epsilon_{\vec{k}}$ is the normalization coefficient. The elements of $\Lambda$ form orthogonal basis in $W^{1,2}(\Gamma)$ and the corresponding physical unknowns can be expressed in a similar form as the series:

\begin{equation}\label{eq:55}
\zeta_1=\sum_{\vec{k}}\widehat{\zeta}_{1,\vec{k}}(t)\frac{2}{\sqrt{L_xL_y}}\epsilon_{\vec{k}}\cos(k_xx)\cos(k_yy),
\end{equation}

\begin{equation}\label{eq:56}
\zeta_2=\sum_{\vec{k}}\widehat{\zeta}_{2,\vec{k}}(t)\frac{2}{\sqrt{L_xL_y}}\epsilon_{\vec{k}}\cos(k_xx)\cos(k_yy),
\end{equation}

\begin{equation}\label{eq:57}
\Phi_1=\sum_{\vec{k}}\widehat{\Phi}_{1,\vec{k}}(t)\frac{2}{\sqrt{L_xL_y}}\epsilon_{\vec{k}}\cos(k_xx)\cos(k_yy),
\end{equation}

\begin{equation}\label{eq:58}
\Phi_3=\sum_{\vec{k}}\widehat{\Phi}_{3,\vec{k}}(t)\frac{2}{\sqrt{L_xL_y}}\epsilon_{\vec{k}}\cos(k_xx)\cos(k_yy),
\end{equation}

\noindent where $\widehat{\zeta}_{\vec{k}}(t)$, $\widehat{\Phi}_{\vec{k}}(t)$ are the spectral wave amplitudes and the perturbed potentials in Fourier space, whereas $\vec{k}=(k_x,k_y)$. Note that the boundary conditions are satisfied in the weak sense only when using the functions (\ref{eq:55})-(\ref{eq:58}). Taking into account the orthogonality properties of the cosine functions the set of wave equations including the boundary conditions can be rewritten (see the Appendix B) in the spectral coefficient space for $\vec{k}$ and $\vec{k}'$ mode interactions:

\begin{eqnarray}\label{eq:59}
\partial_{tt}\widehat{\zeta}_{1,\vec{k}}&+&\gamma_{1}\partial_{t}\widehat{\zeta}_{1,\vec{k}}-R_{c,1}\partial_{tt}\widehat{\zeta}_{2,\vec{k}}+\omega_{1,\vec{k}}^2\widehat{\zeta}_{1,\vec{k}}\nonumber\\
&&=-\sum_{\vec{k}'\geqslant0}\frac{\sigma_1}{4\alpha_1}\epsilon\vec{_k}\epsilon\vec{_{k'}} [(k'_yk_x-k'_xk_y)(\widehat{B}_{k'_x+k_x,k'_y+k_y}-\widehat{B}_{k'_x-k_x,k'_y-k_y})\nonumber\\
&&+(k'_yk_x+k'_xk_y)(\widehat{B}_{k'_x+k_x,k'_y-k_y}-\widehat{B}_{k'_x-k_x,k'_y+k_y})]\widehat{\Phi}_{1,\vec{k}'},
\end{eqnarray}

\begin{eqnarray}\label{eq:60}
\partial_{tt}\widehat{\zeta}_{2,\vec{k}}&+&\gamma_{2}\partial_{t}\widehat{\zeta}_{2,\vec{k}}-R_{c,2}\partial_{tt}\widehat{\zeta}_{1,\vec{k}}+\omega_{2,\vec{k}}^2\widehat{\zeta}_{2,\vec{k}}\nonumber\\
&&=-\sum_{\vec{k}'\geqslant0}\frac{\sigma_3}{4\alpha_2}\cdot\epsilon\vec{_k}\epsilon\vec{_{k'}} [(k'_yk_x-k'_xk_y)(\widehat{B}_{k'_x-k_x,k'_y-k_y}-\widehat{B}_{k'_x+k_x,k'_y+k_y})\nonumber\\
&&+(k'_yk_x+k'_xk_y)(\widehat{B}_{k'_x-k_x,k'_y+k_y}-\widehat{B}_{k'_x+k_x,k'_y-k_y})]\widehat{\Phi}_{3,\vec{k}'},
\end{eqnarray}

\noindent where the new coefficients are defined as:

\begin{equation}\label{eq:61}
\gamma_{1}=\alpha_1^{-1}\rho_1k_{f1}/h_1,
\end{equation}

\begin{equation}\label{eq:62}
\gamma_{2}=\alpha_2^{-1}\rho_3k_{f3}/h_3,
\end{equation}

\begin{equation}\label{eq:63}
R_{c,1}=\alpha_1^{-1}\rho_2/h_2,
\end{equation}

\begin{equation}\label{eq:64}
R_{c,2}=\alpha_2^{-1}\rho_2/h_2.
\end{equation}

\noindent The corresponding uncoupled shallow layer gravity wave frequencies are

\begin{equation}\label{eq:65}
\omega_{1,\vec{k}}^2=R_1\alpha_1^{-1}\vec{k}^2,
\end{equation}

\begin{equation}\label{eq:66}
\omega_{2,\vec{k}}^2=R_2\alpha_2^{-1}\vec{k}^2.
\end{equation}

\noindent The selection of the magnetic field modes in (\ref{eq:59}), (\ref{eq:60}) are obtained from the given magnetic field $B_z^{0}(x,y)$ Fourier expansion in the sine functions:

\begin{equation}\label{eq:67}
\widehat{B}_{k_x,k_y}=\frac{4}{L_xL_y}\int_\Gamma B_z^{0}\sin(k_xx)\sin(k_yy)dxdy,
\end{equation}

\noindent for both positive and negative $(k_x,k_y)=(m\pi/L_x,n\pi/L_y)$. In the particular case of a uniform constant magnetic field $B_z=B_z^{0}=const$ the expansion coefficients are 

\begin{equation}\label{eq:671}
\widehat{B}_{k_x,k_y}=\frac{4B_z^{0}}{mn\pi^2}[1-(-1)^m][1-(-1)^n].
\end{equation}

The set of equations for the potentials in the spectral representation is

\begin{equation}\label{eq:68}
\left(h_{1}h_{2}\vec{k}^2+\sigma_{e,1}\right)\widehat{\Phi}_{1,\vec{k}}=-\frac{j}{\sigma_1}(\widehat{\zeta}_{2,\vec{k}}-\widehat{\zeta}_{1,\vec{k}}),
\end{equation}

\begin{equation}\label{eq:69}
\left(h_{2}h_{3}\vec{k}^2+\sigma_{e,2}\right)\widehat{\Phi}_{3,\vec{k}}=\frac{j}{\sigma_3}(\widehat{\zeta}_{2,\vec{k}}-\widehat{\zeta}_{1,\vec{k}}).
\end{equation}

\noindent The wave equations and the potential equations can be combined by means of the following transformation \citep{Bojarevics94}:

\begin{equation}\label{eq:70}
\widetilde{\zeta}_{1,\vec{k}}=\left(h_1h_2\vec{k}^2+\sigma_{e,1}\right)^{-1/2}\widehat{\zeta}_{1,\vec{k}},
\end{equation} 

\begin{equation}\label{eq:71}
\widetilde{\zeta}_{2,\vec{k}}=\left(h_1h_2\vec{k}^2+\sigma_{e,1}\right)^{-1/2}\widehat{\zeta}_{2,\vec{k}}.
\end{equation}

\noindent The resulting set of wave equations will have the following form suitable for the eigenvalue analysis:

\begin{equation}\label{eq:72}
\partial_{tt}\widetilde{\zeta}_{1,\vec{k}}+\gamma_1\partial_{t}\widetilde{\zeta}_{1,\vec{k}}-R_{c,1}\partial_{tt}\widetilde{\zeta}_{2,\vec{k}}+\omega_{1,\vec{k}}^2\widetilde{\zeta}_{1,\vec{k}}=\sum_{\vec{k}'\geqslant0}\mathsfbi{G}_{1,\vec{k},\vec{k}'}(\widetilde{\zeta}_{1,\vec{k}'}-\widetilde{\zeta}_{2,\vec{k}'}),
\end{equation}

\begin{equation}\label{eq:73}
\partial_{tt}\widetilde{\zeta}_{2,\vec{k}}+\gamma_2\partial_{t}\widetilde{\zeta}_{2,\vec{k}}-R_{c,2}\partial_{tt}\widetilde{\zeta}_{1,\vec{k}}+\omega_{2,\vec{k}}^2\widetilde{\zeta}_{2,\vec{k}}=\sum_{\vec{k}'\geqslant0}\mathsfbi{G}_{2,\vec{k},\vec{k}'}(\widetilde{\zeta}_{2,\vec{k}'}-\widetilde{\zeta}_{1,\vec{k}'}),
\end{equation}

\noindent where the magnetic interaction matrices at the interfaces 1 and 2 respectively are introduced as

\begin{eqnarray}\label{eq:74}
\mathsfbi{G}_{1,\vec{k},\vec{k}'}&=&-\frac{j}{4\alpha_1}\epsilon\vec{_k}\epsilon\vec{_{k'}}[(k'_{y}k_{x}-k'_{x}k_{y})(\widehat{B}_{k'_{x}+k_{x},k'_{y}+k_{y}}-\widehat{B}_{k'_{x}-k_{x},k'_{y}-k_{y}})\nonumber\\
&&+(k'_{y}k_{x}+k'_{x}k_{y})(\widehat{B}_{k'_{x}+k_{x},k'_{y}-k_{y}}-\widehat{B}_{k'_{x}-k_{x},k'_{y}+k_{y}})]\nonumber\\
&&\times\left(h_1h_2\vec{k}^2+\sigma_{e,1}\right)^{-1/2}\left(h_1h_2\vec{k'}^2+\sigma_{e,1}\right)^{-1/2},
\end{eqnarray}

\begin{eqnarray}\label{eq:75}
\mathsfbi{G}_{2,\vec{k},\vec{k}'}&=&-\frac{j}{4\alpha_2}\epsilon\vec{_k}\epsilon\vec{_{k'}}[(k'_{y}k_{x}-k'_{x}k_{y})(\widehat{B}_{k'_{x}-k_{x},k'_{y}-k_{y}}-\widehat{B}_{k'_{x}+k_{x},k'_{y}+k_{y}})\nonumber\\
&&+(k'_{y}k_{x}+k'_{x}k_{y})(\widehat{B}_{k'_{x}-k_{x},k'_{y}+k_{y}}-\widehat{B}_{k'_{x}+k_{x},k'_{y}-k_{y}})]\nonumber\\
&&\times\left(h_1h_2\vec{k}^2+\sigma_{e,1}\right)^{-1/2}\left(h_1h_2\vec{k'}^2+\sigma_{e,1}\right)^{1/2}\left(h_2h_3\vec{k'}^2+\sigma_{e,2}\right)^{-1}.
\end{eqnarray}

\noindent As it can be seen, the $\mathsfbi{G}_{1,\vec{k},\vec{k}'}$ matches the interaction matrix obtained in \citet{Bojarevics94} for the HHC stability description, however $\mathsfbi{G}_{2,\vec{k},\vec{k}'}$ is different and the skew-symmetry for this particular matrix is not retained. The interaction matrices $\mathsfbi{G}_{1,\vec{k},\vec{k}'}$ and $\mathsfbi{G}_{2,\vec{k},\vec{k}'}$ are valid for an arbitrary $B^0_z(x,y)$ expanded according to (\ref{eq:67}).

\subsection{Coupled gravity waves}

Before performing stability analysis of the electro-magnetically caused interactions, let us consider properties of the purely hydrodynamically coupled waves. By neglecting the electro-magnetic and the dissipation terms, equations (\ref{eq:72}) and (\ref{eq:73}) can be solved for the 2 coupled interface gravity wave frequencies:

\begin{equation}\label{eq:755}
\omega^2_{12,\vec{k}}=\frac{-(\omega_{1,\vec{k}}^2+\omega_{2,\vec{k}}^2)\pm[(\omega_{1,\vec{k}}^2-\omega_{2,\vec{k}}^2)^2+4R_{c,1}R_{c,2}\omega_{1,\vec{k}}^2\omega_{2,\vec{k}}^2]^{1/2}}{2(1-R_{c,1}R_{c,2})},
\end{equation}

\noindent where $"+"$ sign stands for lower metal interface and $"-"$ sign for the upper metal interface. The expression is similar to the coupled gravity wave solution in a cylindrical 3 layer system considered in \citet{Weber17}. The physical meaning of the solutions (\ref{eq:755}) is best analysed by solving numerically the wave evolution equations (\ref{eq:72}), (\ref{eq:73}) to inspect specific initial perturbation effects on the two coupled interfaces. For this purpose the 2nd order implicit central finite difference scheme is used to approximate the time derivatives, see Appendix C. The physical variables are reconstructed using (\ref{eq:55}) and (\ref{eq:56}).

\begin{figure}
  \centerline{\includegraphics[width=0.8\textwidth]{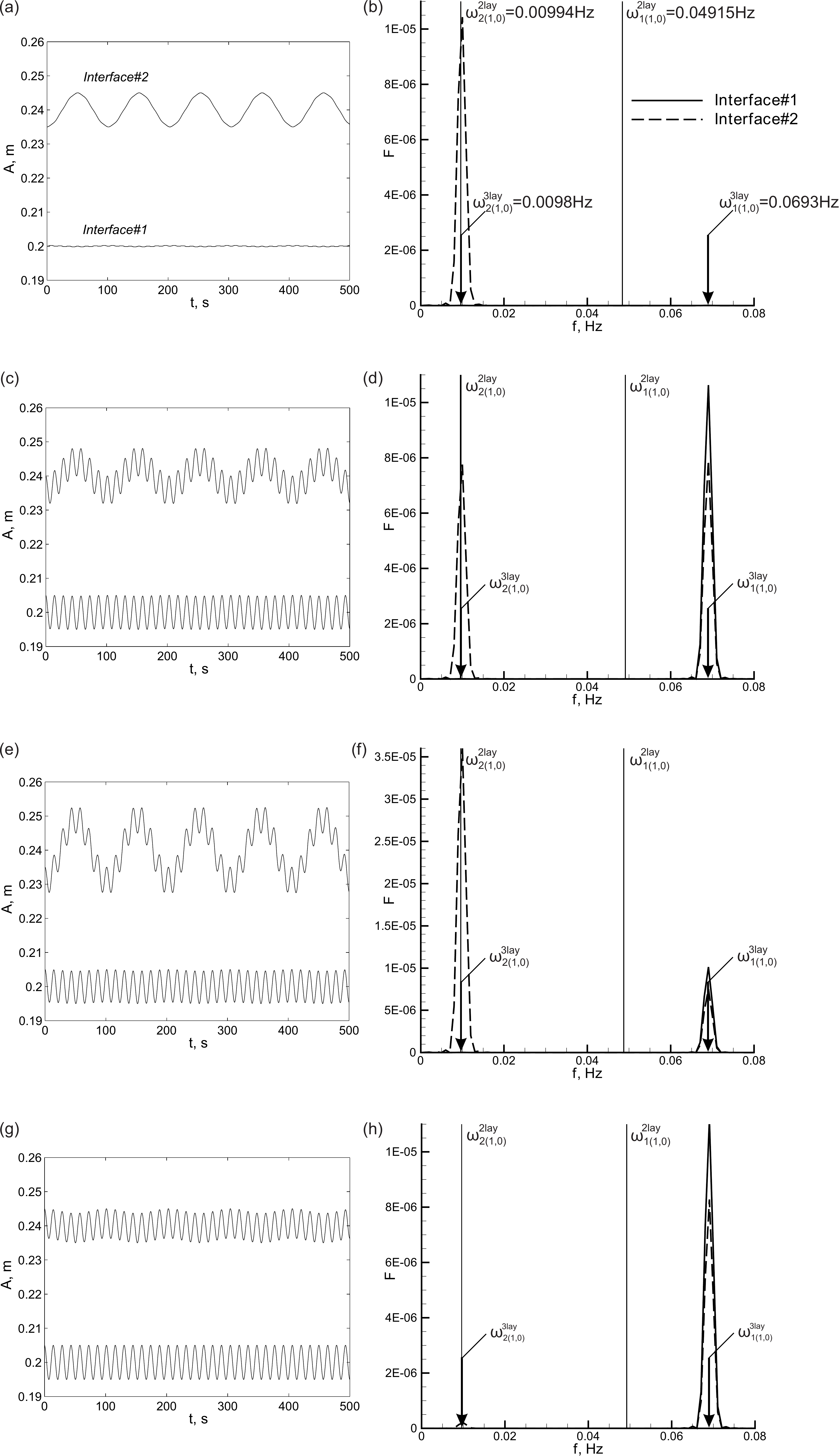}}
  \caption{Solution for the 3 layer coupled gravity waves from the initial value problem with different perturbation types in the Mg$||$Sb battery: the left hand side corresponds to the interface oscillations at the fixed position ($x=0$, $y=0$); the right hand side shows the numerical Fourier power spectra compared to the analytical (\ref{eq:65}), (\ref{eq:66}) and (\ref{eq:755}): (a), (b) Only the top surface is perturbed initially; (c), (d) Only the bottom surface is perturbed; (e), (f) Both surfaces are perturbed asymmetrically at the initial moment; (g), (h) Both surfaces are perturbed symmetrically.}
\label{fig:2}
\end{figure}

The results are shown as interface oscillations at the fixed position in the corner ($x=0$, $y=0$) and the respective Fourier power spectra determined for different initial perturbation types: $(m,n)=\cos(m\pi/L_x)+\cos(n\pi/L_y)$. The first analysed case is for the Mg$\mid$MgCl$_2$-KCl-NaCl$\mid$Sb battery when $\rho_1-\rho_2\gg\rho_2-\rho_3$ (the component physical properties are given in the table \ref{tab:1}), and the large scale rectangular cell with the dimensions: $L_x=8$, $L_y=3.6$ m, and the layer thicknesses: $h_1=0.2$, $h_2=0.04$, $h_3=0.2$ m. The obtained results are summarised in the figure \ref{fig:2}. If only the upper interface is initially perturbed at the amplitude $A=0.005$ m, using the single mode $m=1$, $n=0$, denoted as $(1,0)$,  and the lower interface is initially unperturbed, the initial value problem solution shows that there is only one peak in the spectra, see the figure \ref{fig:2} (a), (b). This indicates that the lower interface remains practically motionless while the upper one is oscillating at the chosen initial perturbation frequency. In this example the 2 layer gravity frequencies: (\ref{eq:65}) and (\ref{eq:66}) can be compared to the 3 layer frequencies defined by (\ref{eq:755}). For the upper interface, 2 and 3 layer approaches match quiet well $\omega^{3lay}_{2(1,0)}\approx\omega^{2lay}_{2(1,0)}$. However this is not the case for the lower interface for which $\omega^{3lay}_{1(1,0)}$ is shifted towards higher frequencies compared to the $\omega^{2lay}_{1(1,0)}$.

In the following example shown in figure \ref{fig:2} (c), (d), when the lower interface is perturbed and the upper is initially unperturbed, a pair of the frequencies are excited in the system. The lower interface oscillates only at the frequency $\omega^{3lay}_{1(1,0)}(\neq\omega^{2lay}_{1(1,0)})$. The spectrum of the upper interface consists of two peaks excited by the lower interface oscillation: $\omega^{3lay}_{1(1,0)}$ and $\omega^{3lay}_{2(1,0)}\approx\omega^{2lay}_{2(1,0)}$. 

\begin{table}
  \begin{center}
\def~{\hphantom{0}}
  \begin{tabular}{lcccc}
   Liquid & $\rho_i$, kg m$^{-3}$  &  $\approx\nu_i$, m$^2$ s$^{-1}$  & $\sigma_i$, S m$^{-1}$ \\[3pt]
    Sb & $6450$   &  $10^{-6}$   & $0.88\times10^6$\\[3pt]
    MgCl$_2$-KCl-NaCl & $1715$   &  $10^{-6}$   & $250$\\[3pt]
     Mg & $1585$   &  $10^{-6}$   & $3.65\times10^6$
  \end{tabular}
  \caption{Material parameters used in numerical examples: density $\rho$, kinematic viscosity $\nu$, conductivity $\sigma$ of the three fluids comprising magnesium-based LMB ($\triangle\rho_1\gg\triangle\rho_2$).}
  \label{tab:1}
  \end{center}
\end{table}

\begin{figure}
  \centerline{\includegraphics[width=0.8\textwidth]{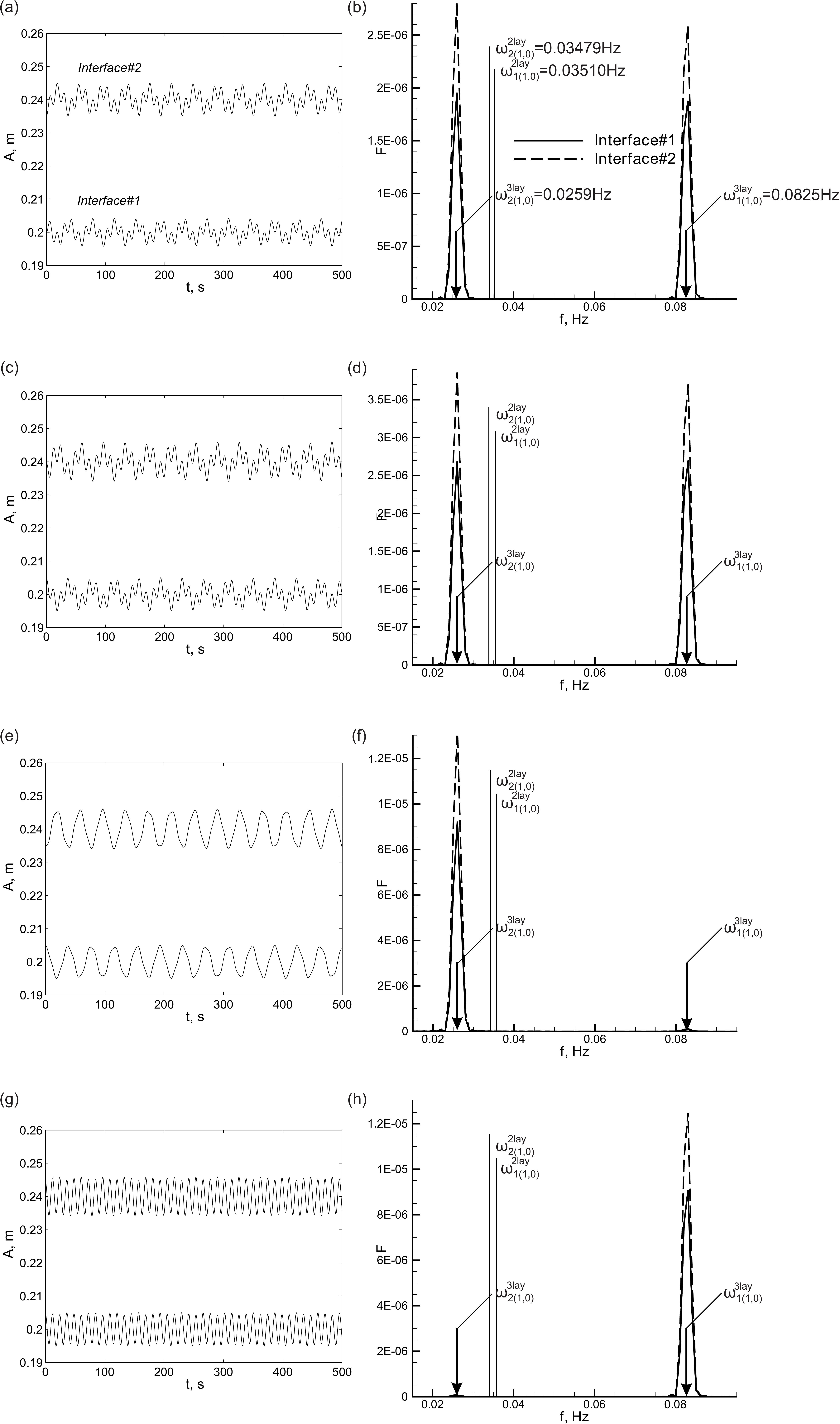}}
  \caption{3 layer coupled gravity waves as initial value problem for different perturbation cases for the Li$||$Te battery: the left hand side corresponds to the interface oscillations at the fixed position ($x=0$, $y=0$); the right hand side shows the Fourier power spectra: (a), (b) Only the top surface is perturbed; (c), (d) Only the bottom surface is perturbed; (e), (f) Both surfaces are perturbed asymmetrically; (g), (h) Both surfaces are perturbed symmetrically.}
\label{fig:3}
\end{figure}

\begin{table}
  \begin{center}
\def~{\hphantom{0}}
  \begin{tabular}{lcccc}
   Liquid & $\rho_i$, kg m$^{-3}$  &  $\nu_i$, m$^2$ s$^{-1}$  & $\sigma_i$, S m$^{-1}$ \\[3pt]
    Te & $5782$   &  $10^{-6}$   & $0.18\times10^6$\\[3pt]
    LiCl-LiF-LiI & $2690$   &  $10^{-6}$   & $250$\\[3pt]
     Li & $489$   &  $10^{-6}$   & $4.17\times10^6$
  \end{tabular}
  \caption{Material parameters used in numerical examples: density $\rho$, kinematic viscosity $\nu$, conductivity $\sigma$ of the three fluids comprising lithium-based LMB ($\triangle\rho_1\approx\triangle\rho_2$).}
  \label{tab:2}
  \end{center}
\end{table}

When both interfaces are initially perturbed in an asymmetric way (in opposite phase) in the (1,0) modes for amplitudes $A=0.005$ m (see the figure \ref{fig:2} (e) and (f)), the qualitative picture of the spectrum is similar to the previous case. The upper one contains a superposition of the two frequencies: $\omega^{3lay}_{1(1,0)}$ and $\omega^{3lay}_{2(1,0)}$, whereas the lower oscillates at a single frequency: $\omega^{3lay}_{1(1,0)}$. 

When the two interfaces are initially perturbed in a symmetric way (in phase) at the respective (1,0) modes, see the figure \ref{fig:2} (g), (h), the wave response is quite different. The upper and lower metal interfaces oscillate at the single frequency: $\omega^{3lay}_{1(1,0)}$.

From the above examples it can be concluded that the coupling of wave dynamics in the considered system is not symmetric. This is due to the significant density difference between the layers $\rho_1-\rho_2\gg\rho_2-\rho_3$ (similar results were obtained in direct numerical simulations by \citet{Weber17}).

The excitation frequency response will be different if $\rho_1-\rho_2\approx\rho_2-\rho_3$. To demonstrate this, an exotic battery case: Li$\mid$LiCl-LiF-LiI$\mid$Te \citep{Kim13} is considered (the component physical properties are given in table \ref{tab:2}). The same system geometry and the perturbation strategy as in the previous examples is used. The obtained results are summarised in figure \ref{fig:3}. If only the upper interface is initially perturbed and the lower interface is initially unperturbed, there are two frequency peaks observed on each of the interfaces, see the figure \ref{fig:3} (a), (b). Both interfaces are set into the motion. Each interface oscillates at $\omega^{3lay}_{1(1,0)}\neq\omega^{2lay}_{1(1,0)}$ and $\omega^{3lay}_{2(1,0)}\neq\omega^{2lay}_{2(1,0)}$. In this case $\omega^{3lay}_{1(1,0)}$ is shifted towards higher frequencies compared to $\omega^{2lay}_{1(1,0)}$, and $\omega^{3lay}_{2(1,0)}$ is shifted towards lower frequencies compared to $\omega^{2lay}_{2(1,0)}$.

In the following example (shown in the figure \ref{fig:3} (c), (d)), when the lower interface is perturbed and the upper is initially unperturbed, the situation is very similar to the previous example. Interfaces are oscillating at the two frequencies $\omega^{3lay}_{1(1,0)}$ and $\omega^{3lay}_{2(1,0)}$. 

When both interfaces are initially perturbed in the asymmetric way (opposite phase), see the figure \ref{fig:3} (e), (f), the qualitative picture of the oscillations and the spectrum changes. The upper and lower interfaces oscillate at the single frequency,$  $ which is $\omega^{3lay}_{2(1,0)}$, while the $\omega^{3lay}_{1(1,0)}$ vanishes from the spectrum.

When both interfaces are initially perturbed in symmetric way (in phase), see the figure \ref{fig:3} (e), (d), the qualitative picture changes again. The upper and lower metal interface oscillates at the frequency $\omega^{3lay}_{1(1,0)}$. In this case  $\omega^{3lay}_{2(1,0)}$ is absent in the spectrum. Similar differences between the symmetric and asymmetric eigenvalue problem solution when $\rho_1-\rho_2\approx\rho_2-\rho_3$ were predicted in \citet{Horstmann17} for the case of cylindrical cell.

\subsection{MHD eigenvalue problem}

Let us proceed now with the stability analysis of the full MHD problem, taking into account the described coupling properties. For this purpose let us assume that the solution form is $\widetilde{\zeta}_{i}\sim e^{\mu t}$, where $\mu$ is representing a set of complex eigenvalues. $\Real(\mu)$ represents the growth rate of instability (when $\Real(\mu)$ is positive the interfacial perturbation $\widetilde{\zeta}$ starts to grow exponentially) and $\Imag(\mu)$ is the electromagnetically modified gravitational wave frequency. The equations (\ref{eq:72}) and (\ref{eq:73}) lead to the following eigenvalue problem:

\begin{equation}\label{eq:76}
(\mathsfbi{A}\mu^2+\mathsfbi{B}\mu+\mathsfbi{C})\cdot\boldsymbol{\zeta}=\boldsymbol{0}.
\end{equation}

\noindent The stability analysis is considerably simplified if restricted to a selected two mode interaction, similarly as in \citep{Bojarevics94}, however accounting for the two interface coupling in the LMB case. Then from (\ref{eq:72}), (\ref{eq:73}) the two mode interaction results in:

\begin{equation}\label{eq:77}
\setlength{\arraycolsep}{0pt}
\renewcommand{\arraystretch}{1.3}
\boldsymbol{\zeta}  = \left[
\begin{array}{c}
  \widetilde{\zeta}_{1,\boldsymbol{k}_1}  \\
  \displaystyle
  \widetilde{\zeta}_{1,\boldsymbol{k}_2}  \\
  \displaystyle
  \widetilde{\zeta}_{2,\boldsymbol{k}_1}  \\
  \displaystyle
  \widetilde{\zeta}_{2,\boldsymbol{k}_2} \\
\end{array}  \right],
\end{equation}

\begin{figure}
  \centerline{\includegraphics[width=1.3\textwidth]{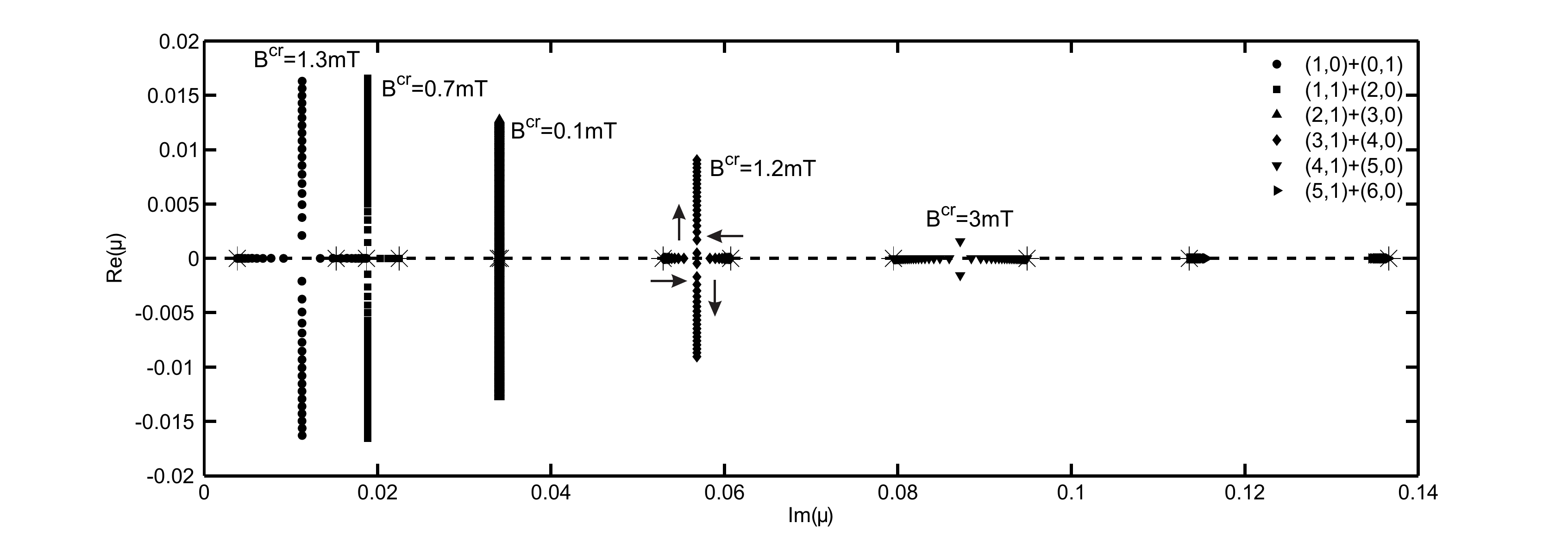}}
  \caption{Eigenvalue analysis for the upper interface: selected eigenvalue couples are moving with the incremental rise of magnetic field $B_{z}=B^{0}_{z}+\Delta B$, where $0\leq B_{z}\leq3$ mT, $\Delta B=0.1$ mT, $L_x/L_y=2.2$; the initial $B_z=0$ position is marked by star signs ($\ast$) and the value of $B^{cr}$ is shown for each interaction.}
\label{fig:4}
\end{figure}

\begin{equation}\label{eq:78}
\setlength{\arraycolsep}{0pt}
\renewcommand{\arraystretch}{1.3}
\mathsfbi{A} = \left[
\begin{array}{cccc}
  1 & 0 & -R_{c,1} & 0  \\
  \displaystyle
  0 & 1 & 0 & -R_{c,1} \\
  \displaystyle
  -R_{c,2} & 0 & 1 & 0  \\
  \displaystyle
  0 &-R_{c,2} & 0 & 1 \\
\end{array}  \right],
\end{equation}

\begin{equation}\label{eq:79}
\setlength{\arraycolsep}{0pt}
\renewcommand{\arraystretch}{1.3}
\mathsfbi{B} = \left[
\begin{array}{cccc}
  \gamma_1 & 0 & 0 & 0  \\
  \displaystyle
  0 & \gamma_1 & 0 & 0 \\
  \displaystyle
  0 & 0 & \gamma_2 & 0  \\
  \displaystyle
  0 & 0 & 0 & \gamma_2 \\
\end{array}  \right],
\end{equation}

\begin{equation}\label{eq:80}
\setlength{\arraycolsep}{0pt}
\renewcommand{\arraystretch}{1.3}
\mathsfbi{C} = \left[
\begin{array}{cccc}
  \omega^2_{1,\vec{k}_1} & \mathsfbi{G}_{1,\vec{k}_1,\vec{k}_2} & 0 & -\mathsfbi{G}_{1,\vec{k}_1,\vec{k}_2}  \\
  \displaystyle
  -\mathsfbi{G}_{1,\vec{k}_1,\vec{k}_2} & \omega^2_{1,\vec{k}_2} & \mathsfbi{G}_{1,\vec{k}_1,\vec{k}_2} & 0 \\
  \displaystyle
  0 & -\mathsfbi{G}_{2,\vec{k}_1,\vec{k}_2} & \omega^2_{3,\vec{k}_1} & \mathsfbi{G}_{2,\vec{k}_1,\vec{k}_2}  \\
  \displaystyle
  -\mathsfbi{G}^T_{2,\vec{k}_1,\vec{k}_2} & 0 & \mathsfbi{G}^T_{2,\vec{k}_1,\vec{k}_2} & \omega^2_{3,\vec{k}_2} \\
\end{array}  \right].
\end{equation}

\noindent Let us consider an example when the applied magnetic field is constant vertical $B_z$ increasing at increments of $\Delta B=0.1$ mT from $0$ to $3$ mT, while the total applied current is fixed: $I=10^5$ A, and the dissipation is neglected (linear friction coefficients $\gamma_1=\gamma_2=0$). The same cell geometry as in the Section 4.2 and the material parameters from the table \ref{tab:1} are used. Selected leading mode interactions for the upper interface are shown in the figure \ref{fig:4}. With the increase of magnetic field the frequencies of the interacting modes are shifted towards each other. When the critical $B^{cr}$ value is reached, the  modes collide, followed by generation of a pair of complex-conjugate modes. One of these gives a positive growth increment that leads to the system destabilization. The results for various mode interactions show that the most dangerous growth rates are for the modes (1,0)+(0,1); (1,1)+(2,0) and (2,1)+(3,0). In the Section 4.6, these findings will be compared with the HHC numerical model \citep{Bojarevics15}. In general, the larger the interacting mode wave number, the larger the critical magnetic field value at which they collide (figure \ref{fig:4}). As expected, for the lower interface all the basic mode interactions in the considered magnetic field range remain stable due to the large density difference of the lower metal and the electrolyte (Mg$||$Sb case). The critical magnetic field is determined by the top interface instability.

\begin{figure}
  \centerline{\includegraphics[width=0.65\textwidth]{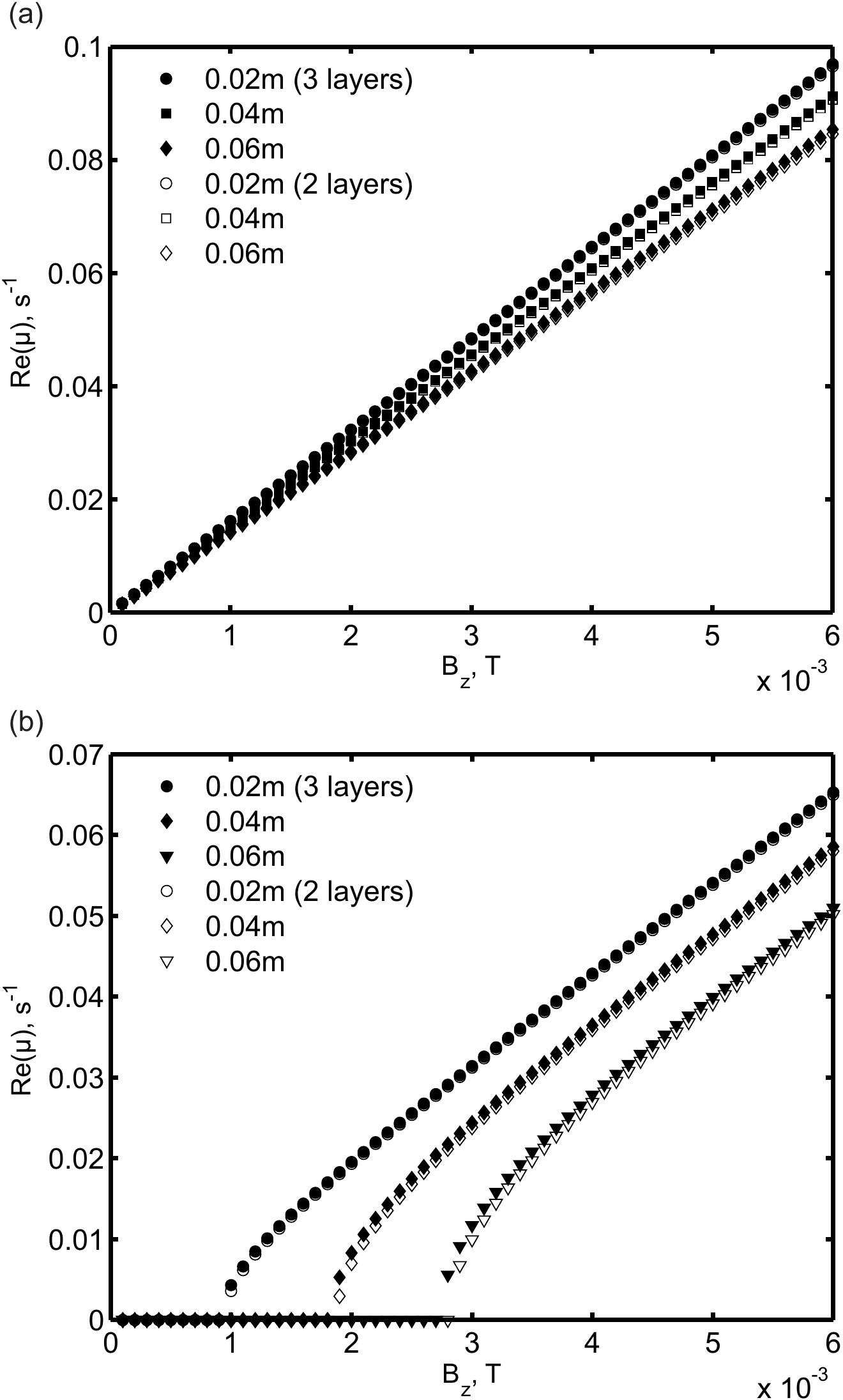}}
  \caption{Comparison of the growth increment dependency on applied magnetic field for the 2 layer and 3 layer models for various values of the electrolyte thickness for $L_x/L_y=1$: (a) The mode $(1,0)+(0,1)$ interaction; (b) $(1,1)+(2,0)$ interaction.}
\label{fig:5}
\end{figure}

When analysing the impact of cell aspect ratio, for instance, $L_x/L_y=1;2;2.5$ etc., the obtained results for interacting modes are in agreement with the results in \citet{Munger08} obtained for the HHC case when the top metal layer stability is dominant in the LMB case.

\subsection{The square cell case}

The previous examples demonstrate that some unperturbed gravity wave mode frequencies are very close in value, however at relatively higher mode orders. In the presence of dissipation these will be damped more rapidly than the leading modes $(1,0)$ and $(0,1)$. In a special case, when the cell aspect ratio $L_x/L_y=1$, the square horizontal section cell is expected to be the most unstable case. The following results are not dependent on the magnitude of the $L_x$, $L_y$ as long as the $\delta$ parameters is sufficiently low to validate the shallow layer approximation. If keeping the same material properties (Mg$||$Sb), total current and the unperturbed current density as in the previous large scale cell examples, the square cell will have the dimensions $L_x=L_y=5.37$ m. Figure \ref{fig:5} shows a comparison of the growth increment dependency on the depth of electrolyte for the linear stability cases of coupled three layers and the two top layers only. As can be seen from the figure \ref{fig:5}, the full three layer model is just marginally different to the two layer model. As previously, we restrict attention to the typical two mode interactions to gain insight to the specific mode interaction mechanisms.

As expected, for the square cell the $(1,0)+(0,1)$ interaction becomes unstable at the smallest magnetic field values for all considered electrolyte depth values. The same result is obtained if reducing the $L_x$ and $L_y$ value down to $0.2$ m and the typical electric current $I\approx130$ A, used in the small scale experimental set-up. In the case of $(1,1)+(2,0)$ mode interaction the stability is retained until a critical magnetic field is reached (at approximately $1$ mT for $h_2=0.02$ m). The stability of the latter interaction increases with the thickness of the electrolyte. The high sensitivity of square cells to the vertical magnetic field is confirmed by \citet{Zikanov17} using the numerical fully coupled nonlinear model based on the shallow layer approximation. 

\subsection{Stability criteria with friction effect}

The effect of bottom friction on the gravity wave damping is analysed in \citet[p. 93]{Landau87} for laminar flow. In reality the friction coefficient values in the equations  (\ref{eq:72}), (\ref{eq:73}) could be significantly higher due to the surface roughness and turbulence generated by the horizontal recirculation flow due to the rotational part of the electromagnetic force in the fluid. The numerical models for aluminium electrolysis cells typically invoke additional turbulence models and empirical values for the bottom friction coefficients, see \citet{Bojarevics15}. The present linear theory can be used to obtain some analytical estimates of the stability criteria for the liquid metal battery MHD waves when the top and bottom friction coefficients are included. 

In the previous sections it was demonstrated that the upper interface stability is the most critical, therefore we will simplify the derivation by restricting to the equation for the $\zeta_2$ interface. This derivation extends the previous results  by adding the effects of friction, while maintaining the electric current redistribution due to the lower metal layer. Based on the assumption that the lower metal is at significantly higher density ($\rho_1\gg\rho_3$, $\zeta_1\rightarrow0$), the wave equation for the upper interface is
 
\begin{eqnarray}\label{eq:81}
\partial_{tt}\widetilde{\zeta}_{\vec{k}}+\gamma\partial_{t}\widetilde{\zeta}_{\vec{k}}+\omega_{\vec{k}}^2\widetilde{\zeta}_{\vec{k}}=\sum_{\vec{k}'\geqslant0}\mathsfbi{G}_{\vec{k},\vec{k}'}\widetilde{\zeta}_{\vec{k}'}.
\end{eqnarray}

\noindent Taking into account that $\left(h_1h_2\vec{k}^2+\sigma_{e,1}\right)\approx\left(h_2h_3\vec{k}^2+\sigma_{e,2}\right)$ (\ref{eq:75}) reduces to:

\begin{eqnarray}\label{eq:82}
\mathsfbi{G}_{\vec{k},\vec{k}'}&=&-\frac{j}{4\alpha_2}\epsilon\vec{_k}\epsilon\vec{_{k'}}[(k'_{y}k_{x}-k'_{x}k_{y})(\widehat{B}_{k'_{x}-k_{x},k'_{y}-k_{y}}-\widehat{B}_{k'_{x}+k_{x},k'_{y}+k_{y}})\nonumber\\
&&+(k'_{y}k_{x}+k'_{x}k_{y})(\widehat{B}_{k'_{x}-k_{x},k'_{y}+k_{y}}-\widehat{B}_{k'_{x}+k_{x},k'_{y}-k_{y}})]\nonumber\\
&&\times\left(h_2h_3\vec{k}^2+\sigma_{e,2}\right)^{-1/2}\left(h_2h_3\vec{k'}^2+\sigma_{e,2}\right)^{-1/2}. 
\end{eqnarray}

The solution of the eigenvalue problem, when two mode interaction stability is considered, can be reduced to a dispersion relation of the 4-th order, that can be written as

\begin{equation}\label{eq:822}
\sum_{n=0}^4 a_n\mu^n=0,
\end{equation}

\noindent  where

\begin{equation}\label{eq:823}
a_0=\omega^2_{\vec{k}_1}\omega^2_{\vec{k}_2}+\vert\mathsfbi{G}_{\vec{k}_1,\vec{k}_2}\vert^2,\quad a_1=\omega^2_{\vec{k}_1}+\omega^2_{\vec{k}_2},\quad a_2=\omega^2_{\vec{k}_1}+\omega^2_{\vec{k}_2}+\gamma^2,\quad a_3=2\gamma,\quad a_4=1.
\end{equation}

\noindent  The explicit solution can be obtained for the selected $\vec{k}_1$ and $\vec{k}_2$ mode interaction:

\begin{equation}\label{eq:83}
\mu=-\frac{\gamma}{2}\pm\left(\Gamma_{\vec{k}_1\vec{k}_2}+(\Delta^2_{\vec{k}_1\vec{k}_2})^{1/2}\right)^{1/2},
\end{equation}

\noindent where

\begin{equation}\label{eq:473}
\Gamma_{\vec{k}_1\vec{k}_2}=\frac{\gamma^2}{4}-\Omega^2_{\vec{k}_1\vec{k}_2},\quad \Omega^2_{\vec{k}_1\vec{k}_2}=\frac{\omega^2_{\vec{k}_1}+\omega^2_{\vec{k}_2}}{2},\quad \Delta^2_{\vec{k}_1\vec{k}_2}=\left(\frac{\omega^2_{\vec{k}_1}-\omega^2_{\vec{k}_2}}{2}\right)^2-\vert\mathsfbi{G}_{\vec{k}_1,\vec{k}_2}\vert^2.
\end{equation} 

\noindent  In the frictionless case ($\gamma=0$) the sufficient condition for the instability is

\begin{equation}\label{eq:84}
\Delta^2_{\vec{k}_1\vec{k}_2}\leqslant0.
\end{equation}

\noindent  If the stability is reached at a finite $\gamma$, then $\Real(\mu)>0$ and (\ref{eq:83}) rewrites as

\begin{equation}\label{eq:85}
\mu=-\frac{\gamma}{2}\pm\left(\Gamma_{\vec{k}_1\vec{k}_2}\pm \mathrm{i}\vert \Delta_{\vec{k}_1\vec{k}_2}\vert\right)^{1/2}.
\end{equation}

\noindent In the case when the system is slightly above the instability threshold ($\vert\Delta_{\vec{k}_1\vec{k}_2}\vert\rightarrow0$), the square root in (\ref{eq:85}) can be expanded in Taylor series, to find the fastest growing mode:

\begin{equation}\label{eq:86}
\mu=-\frac{\gamma}{2}+\Gamma_{\vec{k}_1\vec{k}_2}^{1/2}+\frac{1}{2}\Gamma_{\vec{k}_1\vec{k}_2}^{-1/2}\mathrm{i}\vert \Delta_{\vec{k}_1\vec{k}_2}\vert+\textit{O}(\vert\Delta_{\vec{k}_1\vec{k}_2}\vert^2).
\end{equation}

\noindent For a small friction ($\gamma\rightarrow0$) (\ref{eq:86}) reduces to:

\begin{equation}\label{eq:87}
\mu=-\frac{\gamma}{2}+\mathrm{i}\Omega_{\vec{k}_1\vec{k}_2}+\frac{\vert \Delta_{\vec{k}_1\vec{k}_2}\vert}{2\Omega_{\vec{k}_1\vec{k}_2}}.
\end{equation}

\noindent The system will be unstable if $\Real(\mu)\geq0$, meaning that the friction coefficient

\begin{equation}\label{eq:88}
\gamma\leq\frac{\vert \Delta_{\vec{k}_1\vec{k}_2}\vert}{\Omega_{\vec{k}_1\vec{k}_2}}.
\end{equation}

\noindent The explicit criterion for the instability is

\begin{equation}\label{eq:89}
\gamma\leq\left(\frac{2}{\omega^2_{\vec{k}_1}+\omega^2_{\vec{k}_2}}\right)^{1/2}\left(\vert\mathsfbi{G}_{\vec{k}_1,\vec{k}_2}\vert^2-\left(\frac{\omega^2_{\vec{k}_1}-\omega^2_{\vec{k}_2}}{2}\right)^2\right)^{1/2}.
\end{equation}

\noindent Alternatively the stability condition (\ref{eq:89}) can be derived applying  Routh-Hurwitz (\citet[p. 231]{Gantmacher59}) criterion to (\ref{eq:822}), giving the same expression as (\ref{eq:89}).

\begin{figure}
 \centerline{\includegraphics[width=0.7\textwidth]{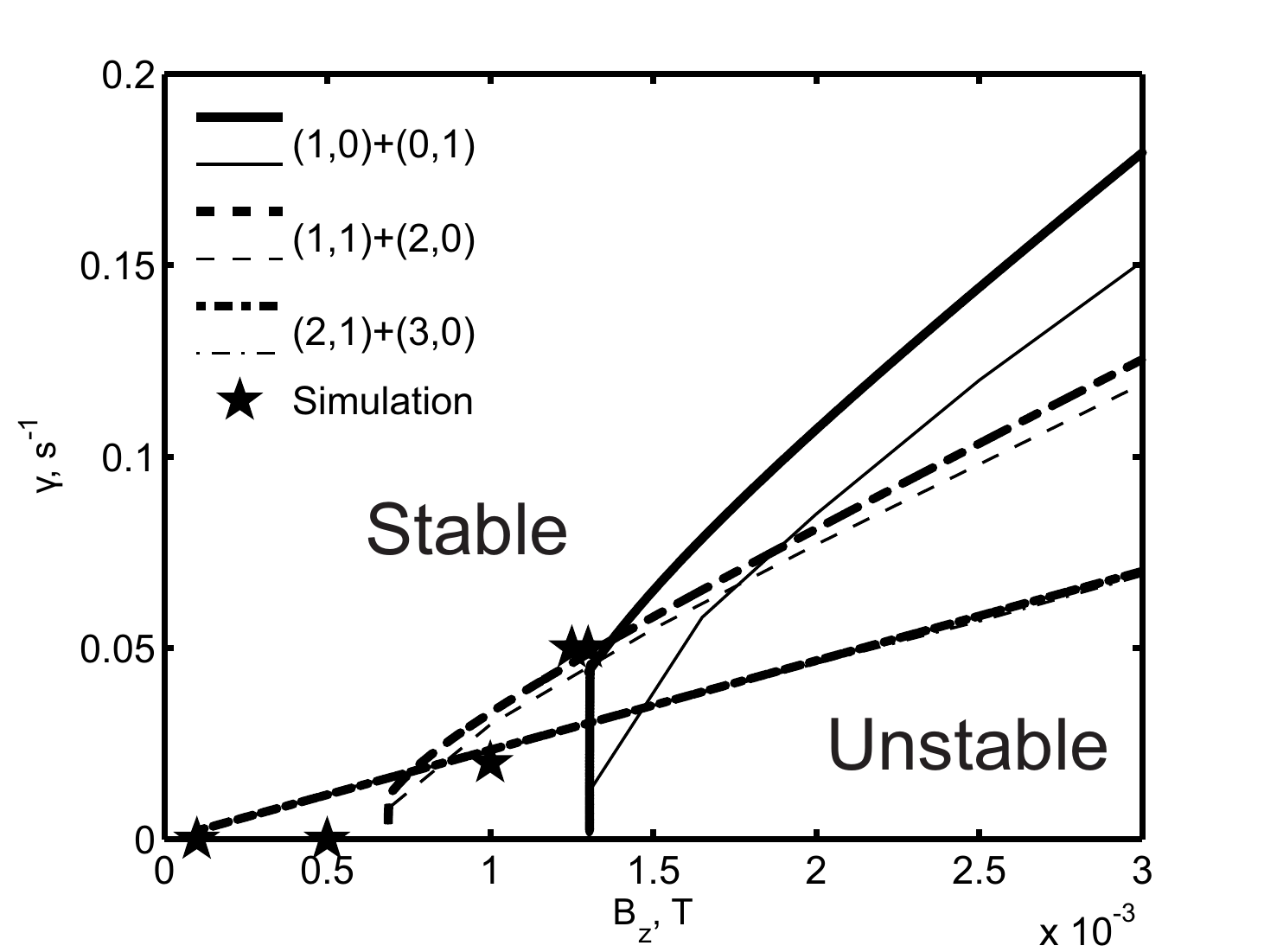}}
  \caption{The critical stabilizing friction dependency on the magnetic field for the two mode interaction: thick lines correspond to the asymptotic relations (\ref{eq:89}), thin lines to (\ref{eq:83}). The symbols $(\star)$ indicate representative numerical test cases shown in the Figure \ref{fig:52}.}
\label{fig:6}
\end{figure}

If the considered modes $\vec{k}_1$ and $\vec{k}_2$ are close enough ($\Imag(\mu_1)\rightarrow\Imag(\mu_2)$) then (\ref{eq:89}) reduces to

\begin{equation}\label{eq:90}
\gamma\leq\left(\frac{2}{\omega^2_{\vec{k}_1}+\omega^2_{\vec{k}_2}}\right)^{1/2}\vert\mathsfbi{G}_{\vec{k}_1,\vec{k}_2}\vert.
\end{equation}

\noindent The results correlating the critical magnetic field $B_z$ and the friction coefficient $\gamma$ are depicted in the figure \ref{fig:6}. The same cell geometry as in Section 4.2 and the material parameters from the table \ref{tab:1} are used for the total electric current $I=10^5$ A. The results show that the asymptotic relation (\ref{eq:89}) gives a good approximation to the general expression (\ref{eq:83}). The relation (\ref{eq:89}) gives the result which is the same as obtained if using the numerical $QZ$ algorithm from the standard, linear algebra software library LAPACK \citep{Andreson99}, except in the case of $(1,0)$, $(0,1)$ interaction with a relatively large gap between the unperturbed gravity frequencies (figure \ref{fig:4}). Note that the discontinuity of stabilizing friction dependency on the magnetic field appears due to the critical magnetic field threshold, below which the system will be always in stable state in the inviscid limit. 

It is instructive to present the general criterion (\ref{eq:89}) in the explicit form for the basic $(1,0)$ and $(0,1)$ mode interaction leading, to the simple stability threshold  criterion:

\begin{equation}\label{eq:900}
\gamma\leq\left[\frac{128j^2B^{02}_z}{(\rho_2/h_2+\rho_3/h_3)(\rho_2-\rho_3)g\pi^6h_2^2h_3^2}\frac{L_x^2L_y^2}{L_x^2+L_y^2}-\frac{(\rho_2-\rho_3)g\pi^2}{2(\rho_2/h_2+\rho_3/h_3)}\frac{(L_y^2-L_x^2)^2}{L_x^2L_y^2(L_x^2+L_y^2)}\right]^{1/2}.
\end{equation}

\noindent In the frictionless case ($\gamma=0$) the stability condition (\ref{eq:900}) by \citet{Bojarevics94} is recovered:

\begin{equation}\label{eq:901}
(\rho_2-\rho_3)g\pi^4h_2h_3\left(\frac{1}{L_y^2}-\frac{1}{L_x^2}\right)=16jB^0_z.
\end{equation}

\noindent The oscillation frequency at the instability onset is defined by (\ref{eq:473}): $\Omega_{\vec{k}_1\vec{k}_2}=\sqrt{(\omega^2_{\vec{k}_1}+\omega^2_{\vec{k}_2})/2}$, which is the rms value of the two interacting gravity modes and can be measured experimentally.

\subsection{Numerical examples} 

\noindent In order to demonstrate the validity of the 2 layer approximation vs 3 layer approximation of the batteries for a selection of materials (the properties given in \citet{Horstmann17}) the fully coupled numerical model was solved using the equations (\ref{eq:27})-(\ref{eq:30}) taking into account the nonlinear wave velocity terms at the right hand side. The electric current redistribution was obtained by solving  (\ref{eq:68})-(\ref{eq:69}). The full 3 layer coupled solutions were compared to the uncoupled 2 layer numerical solutions and the 3 layer respective linear stability results for the same cell geometry as in the Section 4.2, when the total applied current is fixed at $I=10^5$ A while the dissipation is neglected ($\gamma_k=0$). 

The importance of the the ratio of the density differences $(\rho_1-\rho_2)/(\rho_2-\rho_3)$ for the purely hydrodynamic wave coupling was emphasised by \citet{Horstmann17}. The critical magnetic field dependency on this parameter can be analysed using the MHD theory developed in the present paper when neglecting the effects of dissipation. The following approximations were compared: 

\begin{enumerate}
\item linear stability for the 2 layer case,
\item linear stability for 3 layers,
\item decoupled 2 interface simulation,
\item fully coupled 3 layer simulation.
\end{enumerate}

\begin{figure}
  \centerline{\includegraphics[width=1.0\textwidth]{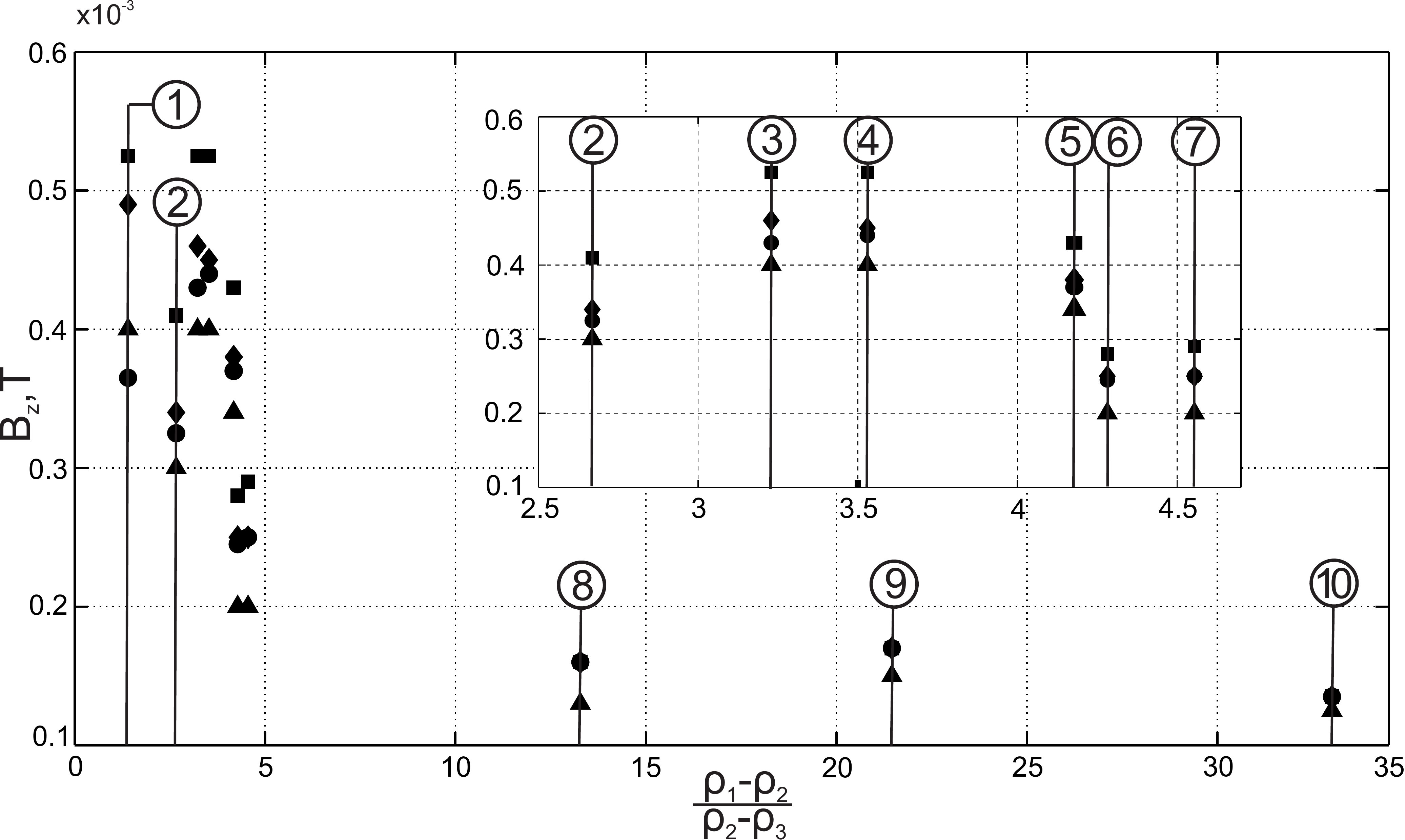}}
  \caption{Critical stability comparison for different battery compositions: ($\blacksquare$) linear stability for the 2 layers, ($\bullet$) linear stability for the 3 layers, ($\blacktriangle$) decoupled 2 interface simulation, ($\blacklozenge$) fully coupled 3 layer simulation.  1) Li$\mid\mid$Te, 2) Na$\mid\mid$Sn, 3) Li$\mid\mid$Bi, 4) Li$\mid\mid$Pb, 5) Na$\mid\mid$Bi, Na$\mid\mid$Pb, 6) Li$\mid\mid$Zn, 7) Li$\mid\mid$Sn, 8) Ca$\mid\mid$Sb, 9) Ca$\mid\mid$Bi, 10) Mg$\mid\mid$Sb.}
\label{fig:55}
\end{figure}

The obtained results are summarized in the figure \ref{fig:55}. Overall a relatively good agreement between all approximations can be seen for the majority of material combinations. The maximum difference for the predicted $B^{cr}$ relative to the fully coupled 3 layer simulation is less than $18\%$. In the range $1<(\rho_1-\rho_2)/(\rho_2-\rho_3)<5$ the 2 layer linear stability overestimates the critical magnetic field, while the decoupled 2 interface simulation underestimates it. In this range the 3 layer linear stability underestimates the system stability if compared to the fully coupled solution. 

The largest difference to fully coupled 3 layer simulation is reached when $(\rho_1-\rho_2)/(\rho_2-\rho_3)\approx1$, which corresponds, e.g. to Li$\mid\mid$Te battery case. For this material combination the 3 layer linear stability predicts the critical magnetic field value $0.365$ mT and the fully coupled approach gives $B^{cr}=0.49$ mT, respectively. 

The lowest critical magnetic field value is found for the Mg$\mid\mid$Sb battery, which emphasises the importance of the interfacial stability for this material combination. In this case both the 2 and 3 layer linear stability predict $B^{cr}=0.135$ mT. The numerical simulation for the decoupled two interfaces gives $0.125$ mT and the fully coupled case results in $0.135$ mT critical value respectively. 

Generally with the decrease of the density ratio, $(\rho_1-\rho_2)/(\rho_2-\rho_3)$, all four approximations predict gradual increase of the critical magnetic field value due to the increased density difference between the upper metal and the electrolyte: $\rho_2-\rho_3$. For the particular case of Na$||$Sn battery a drop of the critical magnetic field is observed due to the lower density difference between the electrolyte and the upper metal ($\rho_2-\rho_3=1619$ kg/m$^3$) if compared to the density differences for Li$||$Te and Li$||$Bi, which are 2201 and 2202 kg/m$^3$ respectively.

\begin{figure}
  \centerline{\includegraphics[width=1.0\textwidth]{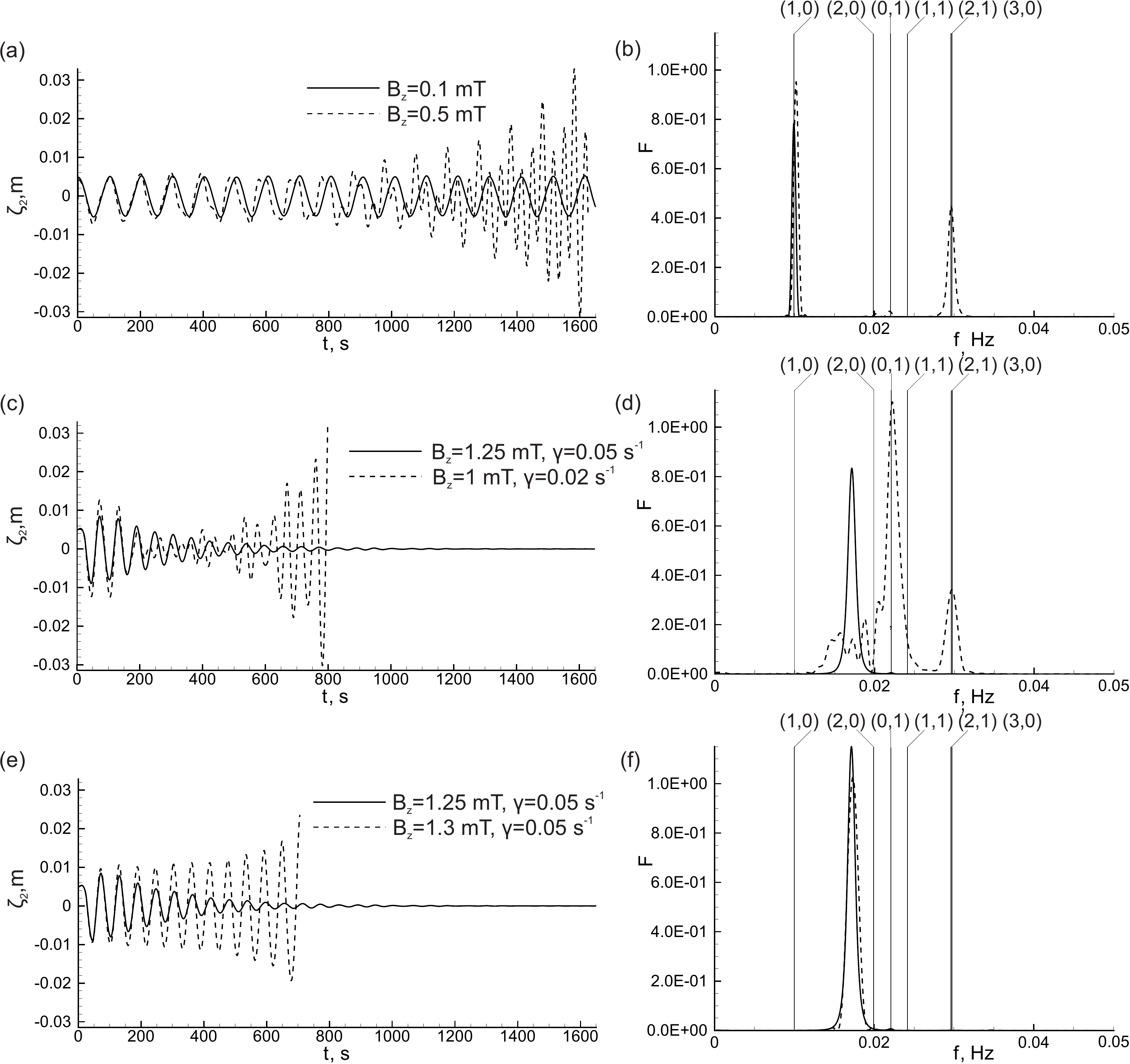}}
  \caption{Numerical results for the top interface oscillation following the initial $(1,0)$ mode perturbation at $A=0.005$ m: (a) oscillation in the frictionless case ($\gamma=0$) for subcritical and overcritical magnetic fields, (b) the power spectra for $\gamma=0$ cases, the black triangles mark the gravity wave frequencies, (c) oscillation in the presence of friction, (d) the power spectra for the two friction coefficients at the marginally stable and unstable cases, (e) oscillation at the higher friction ($\gamma=0.05$ s$^{-1}$) for $B_z$ near the stability limit, (f) the spectral peaks near the stability limit.}
\label{fig:52}
\end{figure}

The full 3 layer numerical solution demonstrates that, independently of the initial perturbation type, the instability of the interfacial motion is generated at the frequency that is located between the two closest orthogonal frequencies. For the most important choice of materials  (Ca$\mid\mid$Sb, Ca$\mid\mid$Bi, Mg$\mid\mid$Sb) onset of instability is determined by the upper interface (figure 6 and 7).

Based on the findings that the top 2 layer model is a good approximation to the LMB stability for the majority of the practically important cases of the material selection, we attempted to compare the previously validated aluminium electrolysis cell numerical models \citep{Bojarevics15} adjusted to the LMB case of the cell geometry given in the Section 4.2, and for the top metal and electrolyte properties given in the table \ref{tab:1}. An additional adjustment was required to include numerically the electric current distribution accounting for the bottom metal layer presence.  The hydrodynamic model permits inclusion of the wave dissipation effects given by the friction coefficient $\gamma$ as in the linear theory. The initial perturbation of the mode $(1,0)$ of amplitude $A=0.005$ m and the total electric current $I=10^5$ A was used in all cases. In the frictionless case $\gamma=0$ at low magnetic field $B_z=0.1$ mT the sloshing wave is continuously oscillating at the same frequency without signs of significant growth or damping (figure \ref{fig:52} a, b). The MHD interaction of the waves becomes unstable at $B_z=0.5$ mT after a large number of oscillation cycles as shown in figure \ref{fig:52} (a). The Fourier transform of the computed time dependent wave amplitude  indicates that the instability sets in due to the dynamic wave transformation resulting in $(2,1)+(3,0)$ mode interaction for this particular cell, figure \ref{fig:52} (b). \citet{Sneyd94} results for the same $(2,1)+(3,0)$ mode interaction in the HHC case predict a very similar value $B^{cr}=0.4$ mT. These results show the advantage of using the coupled fluid dynamic stability analysis over the purely mechanical solid plate model developed by \citet{Davidson98}, $B^{cr}\approx14$ mT, and \citet{Zikanov15}, $B^{cr}\approx11$ mT. The lower stability limit is confirmed independently by the MHD numerical simulations in \citet{Weber17}.

\begin{figure}
  \centerline{\includegraphics[width=1.1\textwidth]{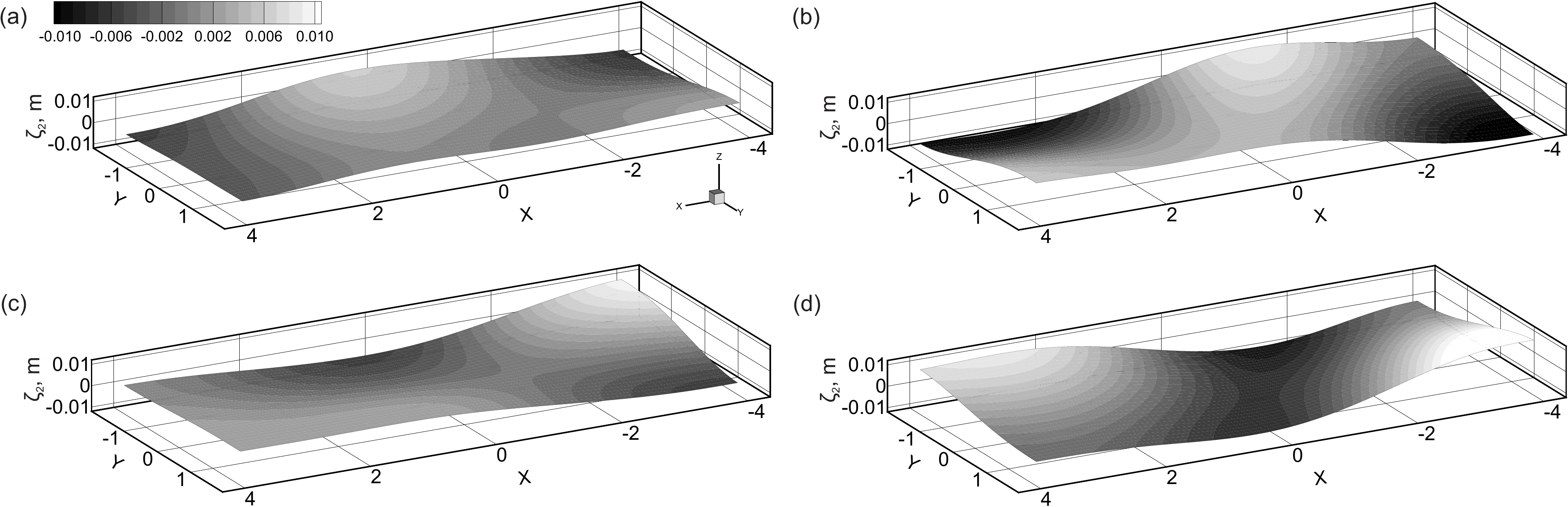}}
  \caption{The computed interface of growing amplitude with the friction $\gamma=0.02$ and $B_z=1$ mT corresponding to Figure \ref{fig:52} (c,d). The frames at $10$ s intervals illustrate the $(1,0)+(0,1)$ and $(2,1)+(3,0)$ mode interactions: (a) $t=635$ s; (b) $t=645$ s; (c) $t=655$ s; (d) $t=665$ s}
\label{fig:53}
\end{figure}

\begin{figure}
  \centerline{\includegraphics[width=1.1\textwidth]{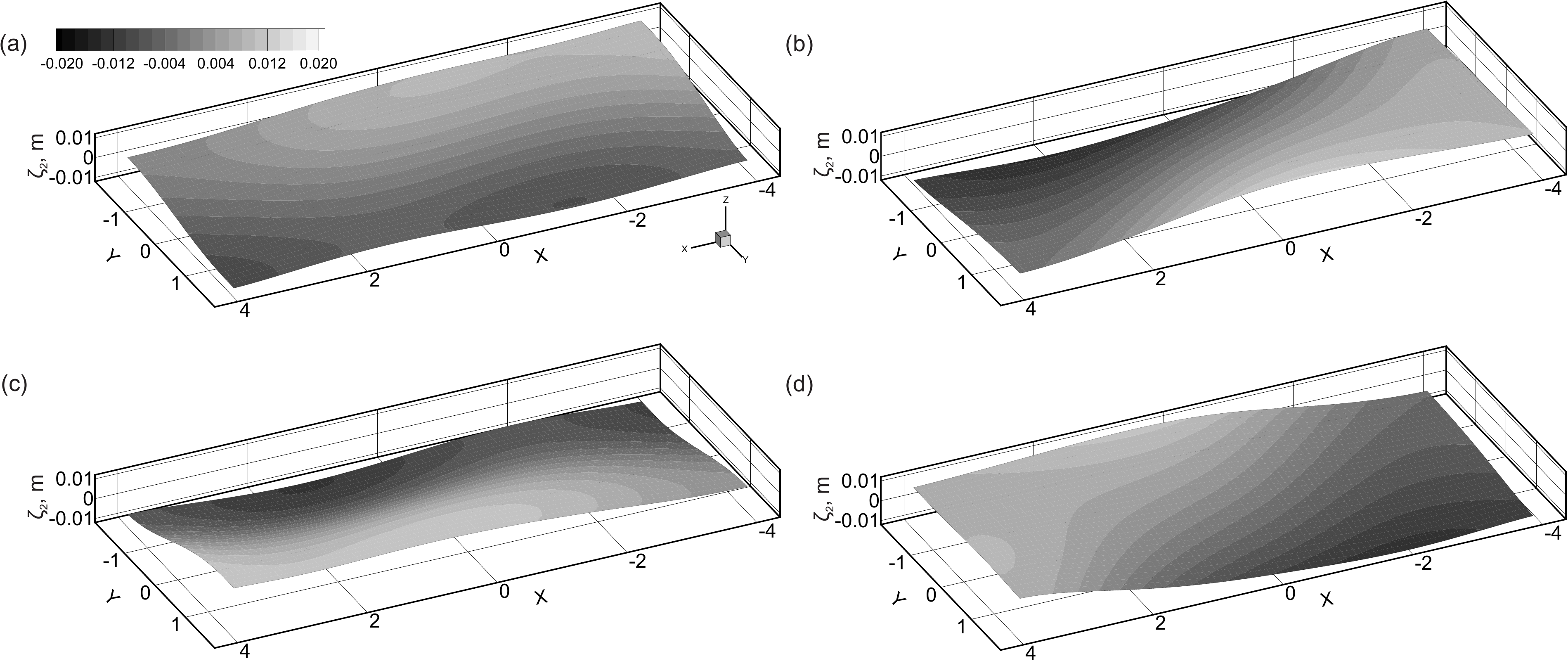}}
  \caption{The computed interface of growing amplitude with $\gamma=0.05$ and $B_z=1.3$ mT corresponding to Figure \ref{fig:52} (e,f) for the $(1,0)+(0,1)$ mode interaction: (a) $t=633$ s; (b) $t=640$ s; (c) $t=660$ s; (d) $t=668$ s}
\label{fig:54}
\end{figure}

After adding the friction coefficient of the empirical value $\gamma=0.05$ s$^{-1}$, which is close to the typical values ($0.02\div0.08$) s$^{-1}$ used for commercial aluminium reduction cells \citep{Zikanov00, Bojarevics15}, the cell becomes unstable at $B_z=1.3$ mT. The oscillation frequency at the instability onset (see the figure \ref{fig:52} (f)) is well described by the two interacting mode rms value $\Omega_{(1,0),(0,1)}=0.017$ Hz (\ref{eq:473}). The transition to the instability is rather sensitive to the $B_z$ value, as can be seen from the figure \ref{fig:52} (c) showing the damped oscillation for $B_z=1.25$ mT. For a lower $B_z$ the damping is dominant, for a higher $B_z$ the growth rate increases, for example, at $B_z=1.5$ mT it takes only 97 seconds for the top interface wave to reach the short circuiting condition at the bottom metal. The typical wave snapshots at the late stage of development are shown in the figures \ref{fig:53} and \ref{fig:54}. The four frames shown at intervals of approximately a quarter of the period ($T_{(2,1)}\approx33.71$ s) are demonstrating a more complex wave rotation pattern than a typical rotating wave along the whole cell perimeter. The instability threshold at $\gamma=0.02$ s$^{-1}$, $B_z=1$ mT, found from the numerical wave evolution simulation, matches the instability onset according to the analytical criterion (\ref{eq:89}) (see the figure \ref{fig:6}).  

The other transition to instability when the longitudinal mode $(1,0)$ interacts to $(0,1)$ transversal mode is reached at a higher friction $\gamma=0.05$ s$^{-1}$ and $B_z=1.3$ mT according to the analytical criterion (\ref{eq:89}) or (\ref{eq:83}), as deduced from the figure \ref{fig:6}. This is confirmed by the direct numerical wave simulation as shown in the figure \ref{fig:52} (e). The corresponding "rotating" wave frames are shown in the figure \ref{fig:54}. The mode $(1,0)$ and $(0,1)$ interaction occurs at the shifted oscillation frequency $\Imag(\mu)=f$ located between the original gravity wave frequencies (figure \ref{fig:52} (f)).

\section{Concluding remarks}\label{sec:filetypes}

The method of regular perturbations using the small depth  $\delta$ and the small amplitude $\varepsilon$  parameters was applied to reduce the full 3 dimensional problem for the electric current distribution and the interface waves in the 3 layer liquid metal battery model.

The linearised equations are solved in the sense of linear stability analysis. For the most important choice of materials in these batteries (high density lower metal and a light metal at the top: Ca$\mid\mid$Sb, Ca$\mid\mid$Bi, Mg$\mid\mid$Sb) the upper interface is the most unstable. The destabilisation mechanism is very similar to behaviour observed in HHC and confirming with the results in \citet{Weber17}, \citet{Bojarevics17} and \citet{Zikanov17}. The upper interface waves can be successfully analysed using a simplified two layer approximation, however accounting for the electric current redistribution in the bottom metal. 

For batteries with comparable density jumps between the layers (Li$\mid\mid$Te, Na$\mid\mid$Sn, Li$\mid\mid$Bi) both interfaces are significantly deformed, and the behaviour is quite different from the HHC \citep{Horstmann17,Zikanov17}. Our results show that the linear stability underestimates the onset of instability compared to the fully coupled 3 layer numerical solution. With the magnetic interaction the asymmetric initial perturbation of the top and bottom layers is always the dominant in generating the instability.

The lowest mode $(1,0)+(0,1)$ interaction is often the most unstable. For square cells it is always the dominant one leading to instability if even the infinitesimal magnetic field is present (this conclusion holds for shallow systems only $\delta\ll1$)  \citep{Bojarevics94, Zikanov17}. The $(1,0)+(0,1)$ interaction can lead to instability in the small size LMBs with $L_x=L_y=0.2$ m.

The dissipation rate is found to be important for practical applications, leading to the faster damping of higher wave modes and the dominance of lower modes. The pure laminar damping \citep{Landau87} $\gamma\approx0.0007$ s$^{-1}$ is insufficient to preclude the instability growth for the large size cells ($L_x=8$, $L_y=3.6$ m, $h_2=0.04$ m) at the typical $B^{cr}\approx0.1$ mT. The instability onset found at $B^{cr}\approx1$ mT and $\gamma=0.02$ s$^{-1}$ is more realistic, and it is very close to typical values of $\gamma$ used in HHC \citep{Zikanov00, Bojarevics15}. The derived new analytical criteria including the damping effects for the stability of two mode interaction are compared against the multiple mode numerical solutions, giving a good match for the instability onset. This indicates that the newly developed analytical criteria including the viscous dissipation effects are equally applicable both for the liquid metal batteries and in the case of aluminium electrolysis cells.\\

\section*{Acknowledgements}

We acknowledge the idea of the two-mode asymptotic interaction to \underline{Michel Romerio}.

\appendix

\section{Determination of expansion coefficients for electric potential}\label{appA. Determination of expansion coefficients for electric potential}

Taking into account the boundary conditions (\ref{eq:32}), (\ref{eq:35}) and (\ref{eq:36}) for the potential and its normal derivative at the corresponding $\varepsilon$ and $\delta$ equal order terms, the following coefficient equalities follow at the lower metal interface $\overline{z}=\overline{H}_1$:

\begin{equation}\label{eq:98}
a_1=A_1=0,
\end{equation}

\begin{equation}\label{eq:99}
c_1=C_1=0,
\end{equation}

\begin{equation}\label{eq:100}
s_1a_2=e_1,
\end{equation}

\begin{equation}\label{eq:101}
s_1A_2=E_1-\overline{H}_{01}\partial_{ii}B_1,
\end{equation}

\begin{equation}\label{eq:102}
a_2\overline{H}_{01}+b_2=b_1,
\end{equation}

\begin{equation}\label{eq:103}
A_2\overline{H}_{01}+a_2\overline{\zeta}_{1}+B_2=B_1.
\end{equation}

\noindent Similarly at he upper metal interface $\overline{z}=\overline{H}_2$:

\begin{equation}\label{eq:104}
a_3=A_3=0,
\end{equation}

\begin{equation}\label{eq:105}
c_3=C_3=0,
\end{equation}

\begin{equation}\label{eq:106}
s_3a_2=e_3,
\end{equation}

\begin{equation}\label{eq:107}
s_3A_2=E_3-\overline{H}_{02}\partial_{ii}B_3,
\end{equation}

\begin{equation}\label{eq:108}
a_2\overline{H}_{02}+b_2=b_3,
\end{equation}

\begin{equation}\label{eq:109}
A_2\overline{H}_{02}+a_2\overline{\zeta}_{2}+B_2=B_3.
\end{equation}

\noindent The supplied current density is given at the bottom and top current collectors, $\overline{z}=\overline{H}_0,\overline{H}_3$:

\begin{equation}\label{eq:110}
\overline{j}_3=\frac{j}{\delta\sigma_3}=e_3,
\end{equation}

\begin{equation}\label{eq:111}
0=E_3-\overline{H}_3\partial_{ii}B_3,
\end{equation}

\begin{equation}\label{eq:112}
\overline{j}_1=\frac{j}{\delta\sigma_1}=e_1,
\end{equation}

\begin{equation}\label{eq:113}
0=E_1-\overline{H}_0\partial_{ii}B_1.
\end{equation}

\noindent From (\ref{eq:98})-(\ref{eq:113}) the unknown coefficients for unperturbed  and perturbed parts can be expressed in terms of $b_1$, $b_3$  and $B_1$, $B_3$ respectively. Then combining (\ref{eq:101}), (\ref{eq:107}), (\ref{eq:111}) and (\ref{eq:113}) leads to the following expression:

\begin{equation}\label{eq:114}
-\frac{s_3}{s_1}(\overline{H}_{01}-\overline{H}_0)\partial_{ii}B_1=(\overline{H}_3-\overline{H}_{02})\partial_{ii}B_3,
\end{equation}

\begin{equation}\label{eq:115}
-\frac{s_3}{s_1}(\overline{H}_{01}-\overline{H}_0)B_1=(\overline{H}_3-\overline{H}_{02})B_3+\phi,
\end{equation}

\noindent where $\phi$ is a function of $x$ and $y$, satisfying $\partial_{ii}\phi=0$. From the boundary conditions (\ref{eq:37}) and (\ref{eq:114}) it follows that on the vertical walls of the cell $\partial_{n}\phi=0$, so that $\phi$ reduces to a constant which is included into $b_1$.  By means of (\ref{eq:101}), (\ref{eq:103}), (\ref{eq:107}), (\ref{eq:109}) with (\ref{eq:114}) the governing set of the equations for the perturbed potentials $\Phi_1$ and $\Phi_3$ (\ref{eq:45})-(\ref{eq:46}) can be derived.

\section{Weak formulation}\label{appB. Weak formulation}

The set of wave equations (\ref{eq:47}), (\ref{eq:48}) with the corresponding boundary conditions (\ref{eq:49}), (\ref{eq:50}) are represented in the weak formulation in the following way:

\begin{eqnarray}\label{eq:94}
&&\int_\Gamma(\partial_{tt}\zeta_1)qd\sigma+\int_\Gamma \gamma_{1}\partial_{t}\zeta_1qd\sigma-\int_\Gamma R_{G,1}\partial_{tt}\zeta_2qd\sigma+\int_\Gamma\frac{(\rho_1-\rho_2)g}{\rho_1/h_1+\rho_2/h_2}(\vec{\nabla}\zeta_1,\vec{\nabla}q)d\sigma\nonumber\\
&&=-\int_\Gamma\frac{\sigma_1}{\rho_1/h_1+\rho_2/h_2}  B_z^{0}(\partial_y\Phi_1\partial_xq-\partial_x\Phi_1\partial_yq)d\sigma,
\end{eqnarray}

\begin{eqnarray}\label{eq:95}
&&\int_\Gamma(\partial_{tt}H_2) qd\sigma+\int_\Gamma  \gamma_{2}\partial_{t}\zeta_2qd\sigma-\int_\Gamma R_{G,2}\partial_{tt}\zeta_1 qd\sigma+\int_\Gamma\frac{(\rho_2-\rho_3)g}{\rho_2/h_2+\rho_3/h_3}(\vec{\nabla}\zeta_2,\vec{\nabla}q)d\sigma\nonumber\\
&&=-\int_\Gamma\frac{\sigma_3}{\rho_2/h_2+\rho_3/h_3}  B_z^{0}(\partial_x\Phi_3\partial_yq-\partial_y\Phi_3\partial_xq)d\sigma,
\end{eqnarray}

\noindent $d\sigma=dxdy$ and the integration is over $\Gamma$. The set of equations for the electric potentials (\ref{eq:45}), (\ref{eq:46}) with the boundary conditions (\ref{eq:37}) give the following weak form:

\begin{equation}\label{eq:96}
\int_\Gamma h_1h_2(\vec{\nabla}\Phi_1,\vec{\nabla}\psi) d\sigma+\int_\Gamma\sigma_{e,1} \Phi_1\psi d\sigma=-\frac{j}{\sigma_1}\int_\Gamma(\zeta_{2}-\zeta_{1})\psi d\sigma,
\end{equation}

\begin{equation}\label{eq:97}
\int_\Gamma h_2h_3(\vec{\nabla}\Phi_3,\vec{\nabla}\psi) d\sigma+\int_\Gamma\sigma_{e,2}\Phi_3\psi d\sigma=\frac{j}{\sigma_3}\int_\Gamma(\zeta_{2}-\zeta_{1})\psi d\sigma.
\end{equation}

\section{Numerical time stepping scheme}\label{appC. Numerical time stepping scheme} 

The set of equations (\ref{eq:72}) and (\ref{eq:73}) was solved using the second order accurate finite difference representation as

\begin{eqnarray}\label{eq:988}
&&\frac{\widehat{\zeta}_{1,\vec{k}}(t_{i+1})-2\widehat{\zeta}_{1,\vec{k}}(t_{i})+\widehat{\zeta}_{1,\vec{k}}(t_{i-1})}{(\triangle t)^2}+\gamma_1\frac{\widehat{\zeta}_{1,\vec{k}}(t_{i+1})-\widehat{\zeta}_{1,\vec{k}}(t_{i-1})}{2\triangle t}\\
&&-R_{c,1}\frac{\widehat{\zeta}_{2,\vec{k}}(t_{i+1})-2\widehat{\zeta}_{2,\vec{k}}(t_{i})+\widehat{\zeta}_{2,\vec{k}}(t_{i-1})}{(\triangle t)^2}+\omega_{1,\vec{k}}^2\frac{\widehat{\zeta}_{1,\vec{k}}(t_{i+1})-\widehat{\zeta}_{1,\vec{k}}(t_{i-1})}{2}=\widetilde{\Xi}_{1,\vec{k}}(t_i),\nonumber
\end{eqnarray}

\begin{eqnarray}\label{eq:999}
&&\frac{\widehat{\zeta}_{2,\vec{k}}(t_{i+1})-2\widehat{\zeta}_{2,\vec{k}}(t_{i})+\widehat{\zeta}_{2,\vec{k}}(t_{i-1})}{(\triangle t)^2}+\gamma_2\frac{\widehat{\zeta}_{2,\vec{k}}(t_{i+1})-\widehat{\zeta}_{2,\vec{k}}(t_{i-1})}{2\triangle t}\\
&&-R_{c,2}\frac{\widehat{\zeta}_{1,\vec{k}}(t_{i+1})-2\widehat{\zeta}_{1,\vec{k}}(t_{i})+\widehat{\zeta}_{1,\vec{k}}(t_{i-1})}{(\triangle t)^2}+\omega_{2,\vec{k}}^2\frac{\widehat{\zeta}_{2,\vec{k}}(t_{i+1})-\widehat{\zeta}_{2,\vec{k}}(t_{i-1})}{2}=\widetilde{\Xi}_{2,\vec{k}}(t_i),\nonumber
\end{eqnarray}

\noindent where $\triangle t$ is the time step. In the numerical examples $\triangle t=0.2$ s was found to be sufficient if comparing to the reduced time step test simulations. The notation $\widetilde{\Xi}_{\vec{k}}$ symbolically represents the combination of the electromagnetic forcing and the non-linear velocity terms.  In the absence of dissipation the numerical solution reproduces the gravity waves corresponding to the analytical solution, see the figures \ref{fig:2} and \ref{fig:3}.

\bibliographystyle{jfm}
\bibliography{Paper_ver3}

\end{document}